\documentclass[fleqn,usenatbib]{mnras}

\usepackage{newtxtext,newtxmath}
\usepackage[T1]{fontenc}
\usepackage{ae,aecompl}
\usepackage[colorinlistoftodos]{todonotes}

\usepackage{graphicx}% Including Figure files
\graphicspath{ {images/} }

\usepackage{amsmath}	% Advanced maths commands
\usepackage{amssymb}	% Extra maths symbols

\newcommand{\mdf}{m_\textrm{\tiny DF}}   % effective dynamical-friction mass   % K17
\newcommand{\tdyn}{\uptau_\textrm{\tiny dyn}}

\title[MBH Binary Analysis with LISA]{Probing Massive Black Hole Binary Populations with LISA}

\author[Michael L. Katz et al.]{
Michael L. Katz,$^{1,2}$\thanks{E-mail: mikekatz04@gmail.com}
Luke Zoltan Kelley,$^{2}$
Fani Dosopoulou,$^{3,4}$
\newauthor Samantha Berry,$^{2,5}$
Laura Blecha,$^{6}$
and Shane L. Larson$^{1,2}$
\\
$^{1}$Department of Physics and Astronomy, Northwestern University,
    Evanston, IL 60208, United States\\
$^{2}$Center for Interdisciplinary Exploration and Research in Astrophysics (CIERA), Evanston, IL, United States\\
$^{3}$Princeton Center for Theoretical Science, Princeton University, Princeton, NJ 08544, USA\\
$^{4}$Department of Astrophysical Sciences, Princeton University, Princeton, NJ 08544, USA\\
$^{5}$Department of Physical Science, Harry S. Truman College , Chicago, IL, United States\\
$^{6}$Physics Department, University of Florida, Gainesville, FL, United States
}

\date{Accepted 2019 October 24. Received 2019 October 15; in original form 2019 August 15}

\pubyear{2018}

\begin{document}
\label{firstpage}
\pagerange{\pageref{firstpage}--\pageref{lastpage}}
\maketitle

\begin{abstract}
ESA and NASA are moving forward with plans to launch LISA around 2034. With data from the Illustris cosmological simulation, we provide analysis of LISA detection rates accompanied by characterization of the merging massive black hole population. Massive black holes of total mass $\sim10^5-10^{10} M_\odot$ are the focus of this study. We evolve Illustris massive black hole mergers, which form at separations on the order of the simulation resolution ($\sim$kpc scales),  through coalescence with two different treatments for the binary massive black hole evolutionary process. The coalescence times of the population, as well as physical properties of the black holes, form a statistical basis for each evolutionary treatment. From these bases, we Monte Carlo synthesize many realizations of the merging massive black hole population to build mock LISA detection catalogs. We analyze how our massive black hole binary evolutionary models affect detection rates and the associated parameter distributions measured by LISA. With our models, we find massive black hole binary detection rates with LISA of $\sim0.5-1$ yr$^{-1}$ for massive black holes with masses greater than $10^5M_\odot$. This should be treated as a lower limit primarily because our massive black hole sample does not include masses below $10^5M_\odot$, which may significantly add to the observed rate. We suggest reasons why we predict lower detection rates compared to much of the literature.
\end{abstract}


%%%%%%%%%%%%%%%%%%%%%%%%%%%%%%%%%%%%%%%%%%%%%%%%%%

%%%%%%%%%%%%%%%%% BODY OF PAPER %%%%%%%%%%%%%%%%%%

\section{Introduction}

With the selection of the Laser Interferometer Space Antenna (LISA; \citeauthor{LISAMissionProposal} \citeyear{LISAMissionProposal}) by the European Space Agency (ESA) for its L3 mission, the massive black hole (MBH) community will gain an important tool for understanding the physics and evolutionary history of MBHs. Following a galaxy merger, the two MBHs from the galactic centers eventually form a bound pair and become a binary. This evolving MBH pair radiates gravitational waves at a range of low frequencies $\sim1\ \text{nHz}-1$ mHz. At the higher end of this range, when the two MBHs are nearer to coalescing, their gravitational wave emission will make them a prime target for the LISA mission. LISA is sensitive to MBH binaries of $\sim10^3-10^9M_\odot$ \citep{Klein2016, LISAMissionProposal, Katz2018} at frequencies from $0.1\ \text{mHz}-10$ mHz corresponding to separations of less than $10^3$ Schwarzschild radii. Signals from these binaries will allow the scientific community to study the origin and evolution of MBHs over cosmic time \citep{GravitationalUniverse, Barausse2015, LISAMissionProposal}. Additionally, MBH binary signals have the potential to reach very high signal-to-noise ratios (SNR, also referred to as $\rho$), unattainable for ground-based gravitational wave detectors \citep{GravitationalUniverse, Barausse2015}. This will allow for high precision measurements of cosmological parameters as well as a greater understanding of fundamental physics \citep{Gair2013, Barausse2015}. 

In this paper, we use the Illustris large-scale cosmological simulations \citep{Vogelsberger2014a, Vogelsberger2014b, Genel2014, Illustris-BHs} to analyze MBH binary populations, their dynamics, and LISA detection prospects. A few papers have provided similar studies. \citet{Blecha2016} performed an analysis on MBH recoil kicks using Illustris data. \citet{PTA-illustris, Kelley2017b} and \citet{Kelley2018} performed a similar analysis to ours in relation to Pulsar Timing Array (PTA) predictions. There have been many papers predicting rates for LISA \citep[e.g.][]{Klein2016, Berti2016, Salcido2016, Bonetti2019}, as well as new predictions for the TianQin gravitational wave observatory, which is similar in construction to LISA \citep{Wang2019}. Many rate prediction papers are built from semi-analytic models (SAM). Most rate predictions with large seeds from SAMs, like those in \citet{Klein2016} and \citet{Berti2016}, use seeds on the order of $10^4M_\odot$. These masses are unresolved in the Illustris simulations; therefore, our overall rate estimates will be lower than those predicted in these papers. Similarly, as delays between galaxy mergers and their central MBH mergers have been included in SAMs, the predicted rates have lowered a bit from $\sim20$ yr$^{-1}$ \citep{Arun2009, Sesana2011} to $\sim8$ yr$^{-1}$ \citep{Klein2016, Berti2016}. However, the inclusion of triple MBH interactions has increased the rate to $\sim20$ yr$^{-1}$ \citep{Bonetti2019}. In addition to using MBH binaries from Illustris, we use improved prescriptions for the MBH binary evolutionary process (delay prescriptions) from \citet{Dosopoulou2017} and \citet{PTA-illustris,Kelley2017b}.

Black holes with masses of $\sim10^6-10^{10}M_\odot$ are usually considered MBHs. These MBHs are believed to exist in the centers of most galaxies of considerable size \citep{Soltan1982, Kormendy1995, Magorrian1998}. This stems from observations of dynamics in the center of other galaxies, as well as our own Milky Way Galaxy. The MBH in the center of the Milky Way Galaxy has been constrained to a mass of $\sim4.1\times10^6M_\odot$ \citep{BoehleGhez2016}. Intermediate mass black holes (IMBH) with masses of $\sim10^2-10^5M_\odot$ have been theorized and  observations of IMBHs have been suggested \citep[e.g.][]{Lin2018, Bellovary2019}, but remain uncertain. However, electromagnetic observations are beginning to find MBHs in between these ranges ($10^5-10^6M_\odot$) in dwarf galaxies \citep{Reines2013, Moran2014, Satyapal2014, Lemons2015, Sartori2015, Pardo2016, Nguyen2018, Nguyen2019}.

%In addition, some groups have identified EM observable candidates of MBH binaries \citep{Graham2015, Bansal2017}.

Within the $\Lambda$CDM paradigm \citep[e.g.][]{White1991}, galactic halos merge. If conditions are right, the MBHs in their centers can form a binary \citep{Begelman1980}. The pair of MBHs will inspiral via interactions with surrounding gas and stars until they reach close enough separations to emit detectable gravitational waves \citep[e.g.][]{Sesana2004, Haiman2009, Sesana2010, Roedig2011, Dosopoulou2017, PTA-illustris, Kelley2017b, Rasskazov2}. The only way to study MBHs to date is through electromagnetic observations, which lead to a potential observational bias in the MBH population measurements \citep{Shen2008, McConnell2013, Shankar2016, Rasskazov2}. LISA will add a strong and independent method for studying these exotic objects. Additionally, the combined measurement of the luminosity distance to a binary with LISA and the electromagnetic measurement of a binary redshift can be used as a ``standard siren'' to measure the Hubble parameter \citep{Schutz1986, Holz2005, BNSstandardsiren}. To perform this measurement, the host galaxy of the binary will generally be needed to get the redshift value. Many groups suggest that some periodic active galactic nuclei could be MBH binaries \citep[e.g.][]{Graham2015a, Charisi2016, Liu2016}.  This type of measurement can help identify the host galaxy and, therefore, the redshift of the binary. However, even without an electromagnetic counterpart, LISA measurements can help constrain various cosmological parameters \citep{Petiteau2011}. In addition to cosmological parameters, EM counterparts of MBH binaries can help illuminate a variety of astrophysical processes, including accretion physics and galaxy evolution \citep{Burke-Spolaor2013, Bogdanovic2015}.

% removed  Notably, LISA will be able to determine MBH parameters to a much higher precision than EM observations, including their masses, spins, and distances. With these precise measurements, the cosmic history of these objects, as well as their interaction with their galactic hosts will be illuminated. In addition to studying populations of MBHs with LISA observations, LISA can make new measurements for fundamental physics and cosmology using gravitational waves from MBH binaries. For example, LISA will be able to measure the spins of MBHs, allowing scientists to probe questions in general relativity, cosmological theory, and the interaction of particle physics with gravity \citep{Gair2013, Barausse2015}. 

%\subsection{Black Hole Formation Channels}\label{sec:formationchannels}

Understanding formation channels of these large MBHs and how they relate to galaxy formation models is an active area of research. Leading theoretical ideas about MBH formation channels largely differ in their considerations for the mass of MBH seeds at early times in the evolution of the Universe. One such scenario involves the direct collapse of pre-galactic halos with a seed mass on the order of $10^4-10^6M_\odot$ at redshifts of $10-20$ \citep{Loeb1994, Begelman2006, Latif2013, Habouzit2016, Ardaneh2018, Dunn2018}. Another scenario involves seeds of $\sim10^3 - 10^4M_\odot$  from runaway cluster collapse \citep{Omukai2008, Devecchi2009, Davies2011, Katz2015}. For smaller seeds, MBH formation channels involve seeds from the collapse of large Population III stars into black holes on the order of $10^2M_\odot$. This would occur at earlier times in cosmic history at redshifts of $20-50$ \citep{Haiman2000, Fryer2001, Heger2003, Volonteri2003, Tanaka2009, Alvarez2009}. Regardless of the formation channel, observations of active galactic nuclei of $\sim10^9M_\odot$ in the centres of galaxies at $z\sim6-7$ indicate MBHs in the universe must have formed quickly on cosmological timescales after the Big Bang, within 1 billion years \citep{Fan2001b, Fan2001a, Fan2006, Mortlock2011}.

It may be that reality is a combination of these three theories. Additionally,  different accretion types and spin values can affect the growth rate of MBHs \citep{Plowman2010, Plowman2011, Sesana2011}. Due to the possible measurement bias of AGNs with EM observations \citep{Lauer2007, Schulze2011}, these possibly conflicting scenarios will benefit greatly from LISA detections at higher redshifts across the entire mass spectrum.

%Generally, SAMs are needed to directly test and compare these different formation scenarios as they are phenomenological models based on observations. SAMs are therefore, unconstrained in the ability to test models. However, they are constrained by their assumptions of evolutionary prescriptions, seeding scenarios, halo density, etc. Numerical simulations are designed with the goal to reproduce observations from first principles. Black hole evolution in numerical simulations is generally constrained by numerical resolution and black hole seeding in galaxies, which is not trivial. We will discuss this further in section \ref{sec:mergerpop}.

%\subsection{Stages of Binary Massive Black Hole Coalescence}\label{sec:stages}

The coalescence of two MBHs occurs after a long dynamical process forcing the two MBHs to decay from $\sim$kpc separations down to merger \citep{Begelman1980, Yu2002, Merritt2005}. Once two galaxies have merged, their central MBHs sink to the center via dynamical friction from interactions with surrounding stars \citep{Chandrasekhar1943, Quinlan1996, Quinlan1997}. Once the two MBHs become gravitationally bound, the dynamical friction formalism breaks down, and individual interactions between singular stars and the binary must be considered. These interactions extract angular momentum from the binary, driving them closer to each other \citep[e.g.][]{Merritt2013, Vasiliev2013}. This regime is the ``stellar hardening'' or ``loss-cone scattering'' regime, which refers to the specific cone in parameter space where stars have to exist in order to extract angular momentum from the binary \citep{Frank1976, Lightman1977}. Following these interactions, the binary becomes close enough to interact with a circumbinary gas disk if gas is present. Generally, the torque from the gas disk is expected to bring the binary closer to coalescence \citep{Haiman2009}. However, there is growing evidence that specific binary parameters and gas disk properties can lead to the gas disk forcing the binary outwards to larger separations \citep{Munoz2019, Moody2019}. After interaction with a gas disk, or if there is little to no gas present, the binary will enter the gravitational wave regime where it will evolve until coalescence. Once the binary enters the gravitational wave regime, its dynamics follow the formalism of \citet{Peters1963} at small separations of $\sim100-1000$ Schwarzschild radii \citep{PTA-illustris}. Several studies have predicted some residual eccentricity when MBH binaries enter the LISA frequency band \citep{Porter2010, Amaro-Seoane2010, Mirza2017, Dosopoulou2017}. However, we will treat all binaries as circular for convenience. This is also conservative as the eccentricity will cause the binaries to merge faster.

We will examine binary lifetime models from \citet{Dosopoulou2017} and \citet{PTA-illustris,Kelley2017b}. In section \ref{sec:lifetimes} we will discuss our models and their specific mathematical approaches to binary lifetime calculations. For this paper, in order to match with cosmological parameters in the Illustris simulation,  we assume a WMAP-9 cosmology with $H_0 = 70.4$ km/s/Mpc, $\Omega_\text{M} = 0.2726$, $\Omega_\text{b}=0.0456$, and $\Omega_\text{vac} = 0.7274$ \citep{Hinshaw2012}.

%\subsection{Previous LISA Predictions}\label{sec:previouspredictions}

\section{Illustris Simulation}\label{sec:Illustris}

The Illustris cosmological simulations are a suite of simulations evolving gas cells, dark matter (DM), star, and MBH particles from $z=137$ to $z=0$ in a cube of side length $106.5$ comoving Mpc. Illustris is based on the moving, unstructured-mesh hydrodynamic code \textit{Arepo} \citep{Springel2010}. Illustris reproduced statistics of large-scale galaxy assembly as well as internal structures of elliptical and spiral galaxies \citep{Vogelsberger2014b}. In particular, we extracted data from Illustris-1, the highest resolution simulation in the suite, with $1820^3$ gas cells and DM particles. DM particles have a mass resolution and typical gravitational-softening length of $\sim6.3\times10^6M_\odot$ and 1.5 kpc, respectively. At redshift zero, there are over $3\times 10^8$ star particles, which have a resolution of $\sim1.3\times10^6M_\odot$ and a typical softening length of 700 pc.

MBHs are implemented in Illustris as massive sink particles. When halos attain a total mass of $7.1\times10^{10}M_\odot$, they are seeded with an MBH of mass $1.42\times10^5M_\odot$ if it does not already have an MBH in it \citep{Illustris-BHs}; the highest density gas cell in the halo is converted to the MBH particle. At this point, the initial dynamical MBH mass will be the same as the gas cell in its previous state. However, the MBH particle is assigned an internal mass of the seed mass, which is tracked from this point \citep{Vogelsberger2013}.

MBH particles grow by Eddington-limited, Bondi-Hoyle accretion from their parent gas cells initially, and then from its nearby gas cells after the dynamical mass of the particle becomes equal to its internal mass. From this point, the dynamical mass and internal MBH mass increase in tandem. The similarity in mass between the MBH particles, gas cells, and star/DM particles would cause the MBH particle to scatter around halos in an unphysical manner without settling down in the center of the halo. Therefore, MBH particles are repositioned to the potential minimum of the host halo at every time step (we refer to this as the ``repositioning algorithm'').

For more overview on the Illustris simulations, see \citet{Vogelsberger2013} and \citet{Torrey2014}. See \citet{Vogelsberger2014a}, \citet{Genel2014}, and \citet{Illustris-BHs} for detailed simulation results and comparisons of the simulations to observations. The initial data sets used in this study were all obtained at \url{www.illustris-project.org} \citep{Nelson2015}. Using this data, we perform more post-processing to create the final datasets used in our analysis. We will further explain this in the following sections.

\subsection{Massive Black Hole Merger Population}\label{sec:mergerpop}

Two MBH particles are merged in the Illustris simulation when they come within a smoothing length of each other ($\sim$kpc). Since these mergers occur at larger scales, we treat this merger event as the formation of the binary, from which we evolve the binary to coalescence with sub-grid models. Following \citet{PTA-illustris}, we adopt the term ``merger'' to indicate this simulation-only process: the combination of two MBH simulation particles into one, indicating the formation of a binary on $\sim$kpc scales. We will refer to the final combination of two realistic MBHs into one as the MBH binary ``coalescence.'' 

Over the course of the simulation, detailed MBH and host-galaxy information is saved in a series of 135 snapshots. Higher time-resolution data was saved for each merger, including the time of the merger and the constituent MBH masses (\citeauthor{Blecha2016} \citeyear{Blecha2016}, \citeauthor{PTA-illustris} \citeyear{PTA-illustris}). We extract detailed properties for all MBHs in the simulation at each snapshot, as well as relevant, global properties for each of their host galaxies from the Illustris ``Group Catalogs'' \citep{Nelson2015}. In addition to the higher time-resolution merger data set, we extract information about host galaxies related to the mergers. For each merger, we locate the host galaxies of the constituent MBHs at the snapshot immediately preceding the merger, as well as the host galaxy of the remnant MBH at the snapshot immediately following the merger. For these galaxies, we not only attained global information, we also gather information about their specific distribution of gas, stellar, and DM constituents. The last data needed for our analysis is the Sublink merger trees \citep{Rodriguez-Gomez2015} to follow galaxies from snapshot to snapshot.

Throughout the cosmic history within Illustris, there are 23,708 MBH merger events. However, a fraction of these mergers are artificial. First, the Friends-of-Friends (FOF) halo finder will occasionally associate two halos as one. When this occurs, the aforementioned repositioning algorithm will force the two MBHs in the centers of each galactic halo to the new potential minimum determined during this misstep by the FOF finder. This causes the two MBHs to merge. After this ``fly-by'' encounter, the two galaxies may separate into two distinct halos as seen by the FOF finder. When this occurs, there will be one galaxy without a central MBH. At this point, a new MBH is seeded in this galaxy causing future artificial mergers to inflate the merger catalog. Similarly, the FOF finder may identify a transient matter overdensity, subsequently seeding a low-mass MBH into the overdensity. This newly seeded MBH is then quickly merged into the MBH in the nearest massive halo due to the repositioning algorithm, once again adding unphysical mergers.

Previously, this was dealt with in \citet{PTA-illustris} and \citet{Blecha2016} using a cut based on mass--only $M_\bullet>10^6M_\odot$ are kept--to exclude low mass MBHs that are overwhelmingly the MBHs involved in these numerical issues. As the authors state, this cut had minimal effect on their predictions for the PTA background, which is dominated by high-mass MBH binaries $M\gtrsim10^8M_\odot$). In a study about LISA, these near-seed mass MBHs play a very important role. Therefore, we designed a post-processing method to avoid removing these small MBHs in the most robust way possible. 

%Within the Illustris simulation, there are certain processes that lead to the numerical issues we strived to address with our advanced extraction analysis.

To handle these issues we start by requiring that all merger constituent MBHs must exist for at least one snapshot prior to the merger, which is always true for MBHs above $10^6 M_\odot$. This removes the MBHs seeded when the FOF finder identifies an overdensity and seeds an unphysical MBH. This is effective in removing these MBHs because the time between the seeding and the merging of the unphysical MBH is less than the duration of one snapshot. We then focus on identifying galaxies that have had their central MBHs removed by the ambiguities related to the FOF finder and the repositioning algorithm. We track the evolution of the galaxy devoid of an MBH as it continues on after its fly-by encounter where it lost its MBH. If this galaxy seeds a new MBH before it merges with another galaxy, we remove this MBH from our catalog, as well as any of its subsequent mergers. 

In addition to filtering the secondary mergers, we analyzed the effects of the premature mergers created by the FOF association of two separate halos. In this process, the removal of the MBH from its galaxy occurs earlier in cosmic time than the actual galactic mergers. This causes the MBH merger to occur at slightly earlier cosmic times. This effect would increase our rate predictions: mergers at earlier cosmic times inflate the number of mergers because the volume of the observer's past light-cone is larger at higher redshifts. However, we believe this effect to be small for two reasons: the binaries involved in this scenario tend to have largely unequal mass ratios, making them increasingly difficult to detect with LISA; and the delay between the unphysical MBH merger and the subsequent galaxy merger is on the order of $10^8$ yrs, which represents a small fraction of the lifetime of the universe as well as a small fraction of the average binary inspiral times predicted from our two binary inspiral models.

%The smaller masses could increase the merger rate as well since larger masses become increasingly harder to detect as their GW frequency approaches the LISA low-frequency band edge. However, this early merger effect usually involves smaller mass MBHs, indicating a slight increase in mass would not substantially change the detectability of the source. 

%After these early mergers, the smaller halo, devoid of an MBH, continues after the encounter where it may seed a new MBH. This new MBH then has the possibility of participating in a future merger, therefore, inflating the merger catalog. To remedy this issue, we follow the evolution of galaxies after their central MBHs are removed. If a new MBH is seeded in this galaxy before a galactic merger occurs, we highlight this as a ``bad'' MBH due to its seeding after the galaxy has lost its true central MBH. We then discard this MBH, as well as its subsequent mergers, from our catalog.
% * <fanid@princeton.edu> 2019-01-07T16:33:25.220Z:
%
% ^.

We will now discuss the post-processing performed on the merger host galaxies as well as the cuts made based on those galaxies.  

\subsection{Host Galaxy Information} \label{sec:hostgalaxies}

To process the host galaxy data, we follow the process of \citet{Blecha2016} and \citet{PTA-illustris}. We need density profiles and velocity dispersions of remnant host galaxies for input into our evolution timescale models. We use the profiles at resolvable scales to extrapolate inward to the centers of the galaxies to infer properties at unresolved scales. Therefore, we confirm each remnant host galaxy is sufficiently resolved for these calculations by requiring the galaxy contain at least 80 DM particles, 80 gas cells, and 300 star particles.

%removed - We require that each constituent MBH exist in an associated galaxy at the snapshot prior to the merger, as well as the remnant MBH after the merger. This cut was satisfied by requiring that each MBH exist for more than one snapshot prior to a merger. Additionally, w

\begin{center}
\begin{figure*}
\includegraphics[width=\textwidth]{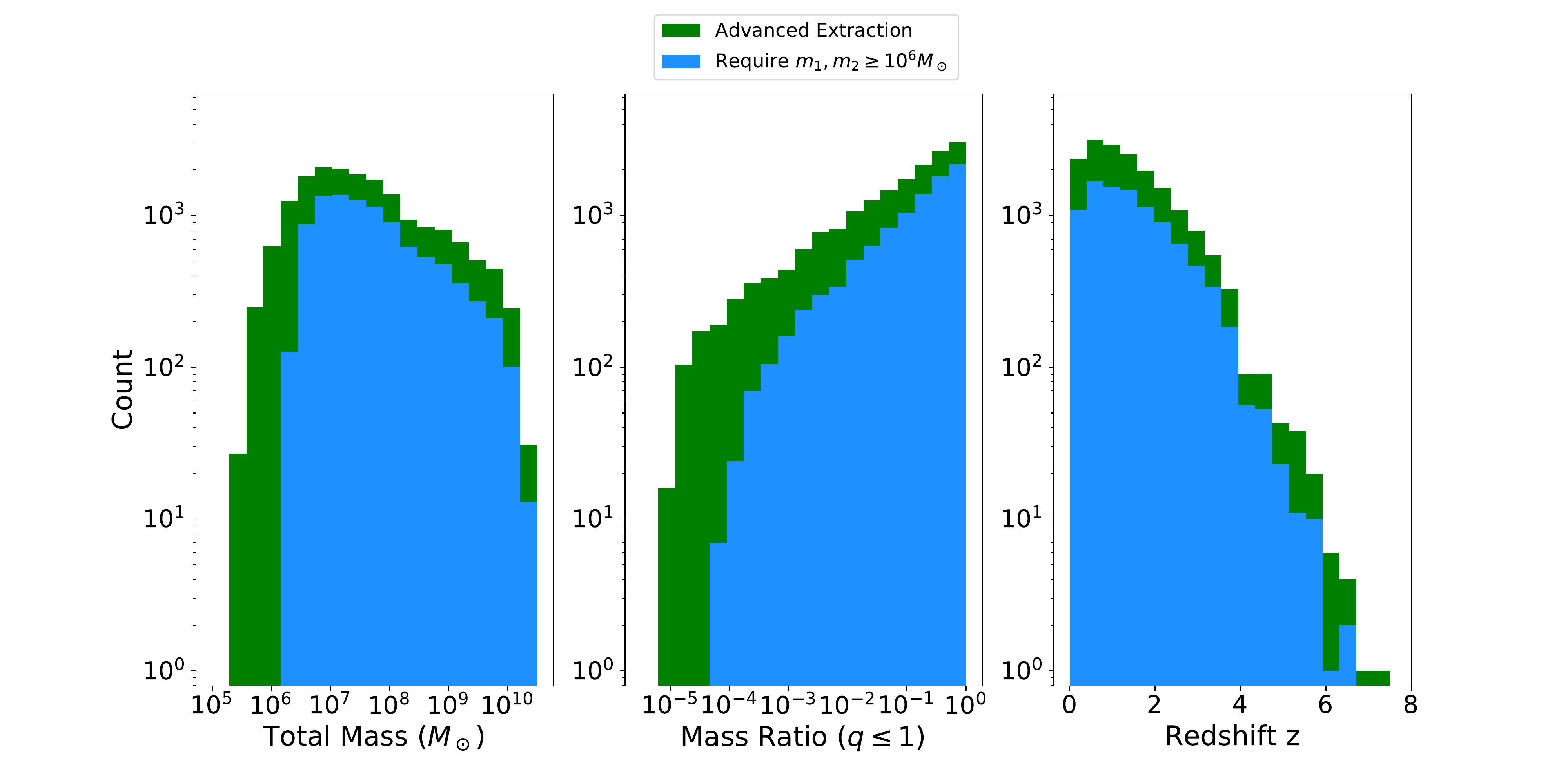}
\caption{Histograms for the main extraction parameters ($M_T$, $q$, and $z$) are shown here. We compare our new advanced extraction (green) to the extraction used previously in \citet{Blecha2016} and \citet{PTA-illustris} requiring $m_1, m_2\geq10^6M_\odot$ (blue). These counts are given after we apply the cuts described in section \ref{sec:Illustris}.}
\label{fig:extract_comparison}
\end{figure*}
\end{center}

\begin{center}
\begin{figure*}
\includegraphics[scale=0.5]{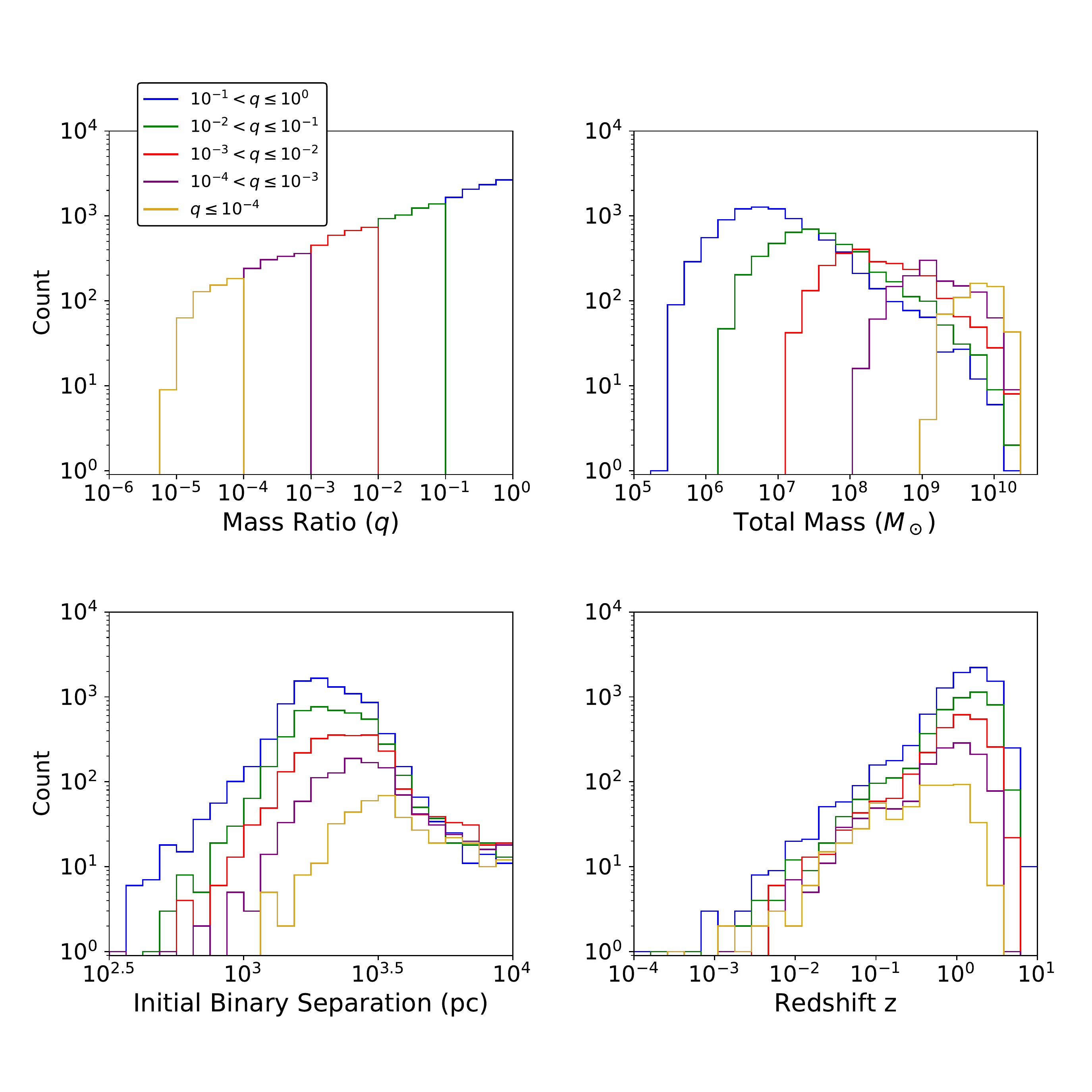}
\caption{Histograms are shown for the binaries that make up our catalog after all of our cuts to the MBH binary population. We group the histograms by mass ratio. The initial separation shown represents the upper limit on the MBH binary separation at binary formation. This is determined from the MBH simulation smoothing length when the MBH particles are merged in the Illustris simulation. Similarly, the redshift here is the redshift at binary formation.}
\label{fig:initial_hists}
\end{figure*}
\end{center}

From these remaining galaxies, we construct spherically averaged, radial density profiles for stars, gas, and DM. We calculate these profiles based on the innermost shells of particles (cells for the gas) surrounding the galactic center. We assume the profile represented by the innermost shells of particles/cells extends inwards to the core of the galaxy. We require at least four particles/cells in each radial bin. The profile is then formed from the innermost 8 bins satisfying the four particle/cell minimum requirement. Binaries were excluded from the final catalogs when fits could not be constructed under these requirements. A graphical representation of this process can be seen in Figure 1 of \citet{PTA-illustris}. For our evolution prescription, we constrain the density profile index to be in between 0.5 and 2.5. The distribution function for the models tested becomes unphysical below an index of 0.5. The upper end of 2.5 is determined based on observed stellar cusps of giant elliptical galaxies. After all of our cuts, we are left with 17,535 of the original 23,708 mergers. We compare this number to 9,270, which is the amount of mergers remaining after the cuts applied in \citet{PTA-illustris}. Therefore, we analyze 8,265 more mergers in this work. The mergers that remain form our final merger catalog. Figure \ref{fig:extract_comparison} compares the main binary parameters resulting from our extraction method to the flat mass cut of $10^6M_\odot$. Figure \ref{fig:initial_hists} shows the main properties of mergers in this final catalog.  

\section{Methods}\label{sec:methods}

\subsection{Binary Lifetime Models}\label{sec:lifetimes}

For the following models, our goal is to calculate the evolutionary timescale from $\sim$kpc scales to coalescence, which is usually of order $\sim$Gyr. We use this timescale to find the coalescence time by adding the evolution time to the time of binary formation (particle merger in Illustris).

The first model we examine has no evolution of the binaries from their particle mergers in the Illustris simulation. We refer to this model as ``ND'' for no delays. In other words, we consider the formation time to equal the coalescence time. This model is our baseline model against which we compare our more detailed models for binary MBH coalescence timescales: it represents the exact prediction from the simulation if the merger process is not modeled below $\sim$kpc scales, which is a common assumption in rate prediction papers. 

In addition to our ND model, we will examine a subset of our ``no delays'' model requiring masses to be greater than $10^6M_\odot$. This allows us to test the difference in our extraction process by comparing the new data set to the old extraction data set similar to the one used in \citet{PTA-illustris}. We will refer to this model as ``ND-6.''

In the following, we will describe detailed models that have been constructed and analyzed in previous papers. Therefore, we give a quick overview of each model. For more information on the DA17 model (section \ref{sec:fanimodel}), please see \citet{Dosopoulou2017}. For the K17 model \mbox{(section \ref{sec:kelleymodel})}, please see \citet{PTA-illustris, Kelley2017b}. 

\subsubsection{DA17 Model}\label{sec:fanimodel}

The equations shown below are taken directly from \citet{Dosopoulou2017}. In what follows, primary, $M$, (secondary, $m$)  will refer to the larger (smaller) MBH. For this model, we are assuming the mergers are gas-poor.

The initial timescale in this model is the large-scale orbital decay from $\sim$kpc scales to a separation equal to the influence radius, $r_{\text{infl}}$, of the primary MBH. $r_{\text{infl}}$ is a shorter length scale than the resolution in the Illustris simulation. Therefore, we use an approximation from \citet{Merritt2009} given by, 
\begin{equation}\label{eq:rinfl}
	r_{\text{infl}} = 10.8\left(\frac{M}{10^8M_\odot}\right)\left(\frac{\sigma}{200\text{ km s}^{-1}}\right)^{-2}\text{pc},
\end{equation}
where $\sigma$ is the three dimensional stellar velocity dispersion of the primary galaxy. Modeling the primary host galaxy as a singular isothermal sphere, \citet{BinneyTremaineGalacticDynamics} show this decay timescale is given by,
\begin{equation}\label{eq:binneytremaine}
	T_{\star}^{\text{bare}} = 17 \frac{6.6}{\ln{\Lambda}}  \left(\frac{R_e}{10\text{ kpc}}\right)^2  \left(\frac{\sigma}{300\text{ km s}^{-1}}\right)  \left(\frac{10^8M_\odot}{m}\right) \text{Gyr},
\end{equation}
where $\ln{\Lambda}$ is the Coulomb logarithm and $R_e$ is the effective radius of the primary galaxy. However, \citet{Dosopoulou2017} modify this formalism to include the effect of the secondary MBH remaining embedded in a core of stars from the secondary galaxy. This added mass causes the system to sink faster towards the primary galaxy's center, therefore, taking less time than is predicted by Equation \ref{eq:binneytremaine}. Assuming the mass of stars bound to the secondary to be a constant proportionality of $10^3m$ \citep{Merritt2001}, this large-scale decay timescale becomes,
\begin{equation}\label{eq:largescaledecay1}
	T_{\star,1}^{\text{gx}} = 0.06 \frac{2}{\ln{\Lambda'}}  \left(\frac{R_e}{10\text{ kpc}}\right)^2  \left(\frac{\sigma}{300\text{ km s}^{-1}}\right)  \left(\frac{10^8M_\odot}{m}\right) \text{Gyr},
\end{equation}
where $\Lambda'=2^{3/2}\sigma/\sigma_s$ with $\sigma_s$ representing the stellar velocity dispersion of the secondary galaxy. This equation, however, does not include tidal stripping of stars from the secondary galaxy by the primary galaxy. Including this effect, the large-scale decay timescale is given by,
\begin{equation}\label{eq:largescaledecay2}
	T_{\star,2}^{\text{gx}} = 0.15 \frac{2}{\ln{\Lambda'}}  \left(\frac{R_e}{10\text{ kpc}}\right)  \left(\frac{\sigma}{300\text{ km s}^{-1}}\right)^2  \left(\frac{100 \text{ km s}^{-1}}{\sigma_s}\right) \text{Gyr}.	
\end{equation}
Using Equations \ref{eq:largescaledecay1} and \ref{eq:largescaledecay2}, we can approximate the large-scale decay timescale as
\begin{equation}\label{eq:largescaledecaymax}
	T_\star = \text{max}\left(T_{\star,1}^{\text{gx}}, T_{\star,2}^{\text{gx}}\right).
\end{equation}
During the large-scale decay, the stellar velocity distribution is treated as Maxwellian. Once the secondary MBH reaches $r_{\text{infl}}$ of the larger MBH, this assumption no longer holds because the potential is dominated by the central MBH. Therefore, the stellar velocity distribution is treated according to Equation 20 in \citet{Dosopoulou2017}. We refer to this next regime as the ``dynamical friction regime'' to match the conventions of the original paper. The timescale for the binary to decay to a shorter separation $r=\chi r_{\text{infl}}$ ($\chi<<1$) is given by,
\begin{equation}\label{eq:tdotbare}
\begin{aligned}
	T_\bullet^{\text{bare}}=&1.5\times10^7 \frac{\left[\ln{\Lambda} \alpha + \beta + \delta\right]^{-1}}{\left(3/2-\gamma\right)\left(3-\gamma\right)}\left(\chi^{\gamma-3/2}-1\right) \\ 
    &\times\left(\frac{M}{3\times10^9M_\odot}\right)^{1/2}\left(\frac{m}{10^8M\odot}\right)^{-1}\left(\frac{r_{\text{infl}}}{300\text{ pc}}\right)^{3/2}\text{ yr},
\end{aligned}
\end{equation}
where $\gamma$ is the power law exponent in the stellar density profile, $\rho(r) = \rho_0\left(r/r_{\text{infl}}\right)^{-\gamma}$; $\alpha$, $\beta$, and $\delta$ are calculated from Equations 17-19 in \citet{Dosopoulou2017}.\footnote{In these equations, we assume a circular orbit setting $\xi=1$ (see the paper for more details). However, the dynamical friction decay timescale is not greatly affected by the orbital eccentricity.} If we include the stars bound to the secondary, as in Equation \ref{eq:largescaledecay2}, this timescale becomes,
\begin{equation}\label{eq:tdotgx}
\begin{aligned}
	T_\bullet^{\text{gx}}=&1.2\times10^7 \frac{\left[\ln{\Lambda} \alpha + \beta + \delta\right]^{-1}}{\left(3-\gamma\right)^2}\left(\chi^{\gamma-3}-1\right) \\ 
    &\times\left(\frac{M}{3\times10^9M_\odot}\right)\left(\frac{100 \text{ km s}^{-1}}{\sigma_s}\right)^{3}\text{ yr}.
\end{aligned}
\end{equation}
Similar to Equation \ref{eq:largescaledecaymax}, we find $T_\bullet$ with, 
\begin{equation}\label{eq:tdotmin}
	T_\bullet = \text{min}\left(T_\bullet^{\text{bare}},T_\bullet^{\text{gx}}\right).
\end{equation}
We use Equation \ref{eq:tdotmin} to evolve the binary down to the hardening radius, $a_h$, given by \citep{Merritt2013},
\begin{equation}\label{eq:a_h}
	a_h\approx36\frac{q}{\left(1+q\right)^2}\frac{M+m}{3\times10^9M_\odot}\left(\frac{\sigma}{300\text{ km s}^{-1}}\right)^{-2} \text{ pc}, 
\end{equation}
where $q$ is the mass ratio ($q\leq1$). Therefore, we set \mbox{$\chi=a_h/r_\text{infl}$}.

The final phase in the DA17 model is the ``hardening phase'', as it includes the effect of gravitational radiation. With dry mergers, remnant galaxies after a merger are expected to be triaxial. In this configuration, an efficient hardening of the binary is exposed to a full and consistently refilling loss-cone (\citeauthor{Khan2011} \citeyear{Khan2011}, \citeauthor{Vasiliev2014} \citeyear{Vasiliev2014}). From $a_h$ until coalescence, including the gravitational wave regime, the timescale is given by \citep{Vasiliev2015},
\begin{equation}\label{eq:thardening}
\begin{aligned}
	T_\text{h,GW} \approx & 1.2\times 10^9\left(\frac{r_\text{infl}}{300 \text{ pc}}\right)^{\frac{10+4\psi}{5+\psi}}\left(\frac{M+m}{3\times10^9 M_\odot}\right)^{\frac{-5-3\psi}{5+\psi}} \\
    \times\ &\phi^{-\frac{4}{5+\psi}}\left(\frac{4q}{\left(1+q\right)^2}\right)^{\frac{3\psi-1}{5+\psi}} \text{ yr}, 
\end{aligned}
\end{equation}
where $\phi=0.4$ and $\psi=0.3$ are triaxial parameters estimated from Monte Carlo simulations in \citet{Vasiliev2015}. In Equation \ref{eq:thardening}, we left out the eccentricity factor as it is unity because we are assuming circularity. Gravitational radiation takes over at $a_\text{GW}$, determined from the ratio \citep{Vasiliev2015},
\begin{equation}\label{eq:ahaGW}
	\frac{a_h}{a_\text{GW}}\approx 55 \left(\frac{r_\text{infl}}{30\text{ pc}}\right)^{5/10}\left(\frac{M+m}{10^8M_\odot}\right)^{-5/10}\left(\frac{4q}{\left(1+q\right)^2}\right)^{4/5},
\end{equation}
where we have once again left out the eccentricity factor as it is equal to unity. 
For $q\approx10^{-3}$, $a_h$ is less than $a_\text{GW}$. For these binaries, we use the inspiral time for a circular binary due to only gravitational radiation according to \citep{Peters1963},
\begin{equation}\label{eq:gw}
	T_\text{GW} = 2.3\times10^6\left(\frac{a_h}{10^{-3}\text{pc}}\right) \left(\frac{10^8M_\odot}{M}\right)^2 \left(\frac{10^5M_\odot}{m}\right) \left(\frac{1}{1+q}\right) \text{ yr}.
\end{equation}
Therefore, our final timescale, $T_\text{final}$ is given by,
\begin{equation}\label{eq:Tfinal}
	T_\text{final}=
    \begin{cases}
    T_\text{h,GW}, & \text{if}\ q\geq10^{-3} \\
    T_\text{GW}, & \text{if}\ q<10^{-3}.
    \end{cases}
\end{equation}

The DA17 model's final coalescence timescale, $t_\text{coal}$, is therefore given by
\begin{equation}\label{eq:Dosopouloutimescale}
	t_\text{coal}=T_\star+T_\bullet+T_\text{final}.
\end{equation}

\subsubsection{K17 Model}\label{sec:kelleymodel}

This section introduces the K17 model, described in detail in \citet{PTA-illustris} and \citet{Kelley2017b}.  Binaries are numerically integrated from their formation at large-separations until their eventual coalescence.  Dynamical friction is implemented following \citet{Chandrasekhar1943}, where the deceleration is given as,
\begin{equation}
    \frac{dv}{dt}\bigg|_\textrm{DF} = -\frac{2\pi G^2 (M+\mdf)\rho}{v^2} \ln \Lambda_c,
\end{equation}
where the relative velocity is taken to be the maximum of the orbital velocity and stellar velocity dispersion, i.e.~\mbox{$v = \max\left(v_\textrm{orb}, \sigma\right)$}, $\rho$ is the total mass-density, and the Coulomb logarithm is set to $\ln \Lambda_c = 15$.  The effective mass of the secondary, $\mdf$, assumes that the mass of the secondary host galaxy is stripped over the course of a dynamical time, i.e.,
\begin{equation}
    \mdf = m \left(\frac{m + m_\textrm{\tiny host}}{m}\right)^{1 - t/\tdyn}.
\end{equation}
The initial host galaxy mass is measured from Illustris in the snapshot preceding the merger event.  Once the binary shrinks below the ``loss-cone radius'' \citep{Begelman1980}, the hardening rate is calculated following the stellar-scattering prescription from \citet{sesana2006}, % and \citet{sesana2010}
\begin{equation}
    \frac{da}{dt}\bigg|_\textrm{SS} = - \frac{G \rho}{\sigma} a^2 \, H,
\end{equation}
where $a$ is the semi-major axis of the binary which is being integrated, and $H$ is a dimensionless coefficient calculated from numerical scattering experiments.

The accretion rate calculated in Illustris provides an estimate of the presence of circumbinary gas.  To model the energy extraction from this material, we assume the gas settles into a geometrically-thin alpha-disk \citep{Shakura1973} with different regions corresponding to the dominant components of pressure (radiation vs.~thermal) and opacity (Thomson vs.~free-free), following \citet{Shapiro1986}.  The hardening rates in each regime are calculated in \citet{Haiman2009}, as a function of disk surface-density and binary mass-ratio.  Numerical simulations have found these analytic prescriptions to be quite accurate over the parameter ranges studied \citep{Tang2017,Fontecilla2019}.  We assume these disks extend out to the radius at which they become Toomre unstable.  GW energy extraction is implemented at all radii, following \citet{Peters1963}.

\subsection{Determining Detectability}\label{sec:detectability}
\subsubsection{Characteristic Strain}\label{sec:charstrain}

MBH binaries provide a variety of signals measurable by LISA since their chirp evolution in the frequency domain occurs near the low-frequency band edge of the LISA sensitivity curve. Binaries with $\sim10^5-10^7M_\odot$ total mass will provide a measurable inspiral, merger, and ringdown leading to very loud signals even out to the cosmic horizon \citep{LISAMissionProposal}.

The binary inspiral is the initial stage of binary black hole coalescence when the two MBHs orbit one-another at separations greater than the innermost stable circular orbit (ISCO; $R=6GM/c^2$). At these separations, the orbit is usually treated with a post-Newtonian formalism.

The merger stage follows the binary inspiral with a highly non-linear relativistic process. This process continues until the MBHs have contacted each other to form a single event horizon, leading to ringdown. The dominant mode of the ringdown spectrum is expected to be the $l=m=2$ quasinormal mode. Deviations from general relativity can be measured if LISA can detect subdominant modes in the ringdown spectrum. This process is referred to as the so-called ``black hole spectroscopy'' \citep{Berti2006, Berti2016, Baibhav2018a, Baibhav2018b}.

We use the characteristic strain, $h_c$, to model the binary signal which accounts for the time the binary spends in each frequency bin \citep{Finn2000}. The characteristic strain is given by \citep{Moore2015},
\begin{equation}
	h_c^2 = 4f^2\left|\tilde{h}(f)\right|^2,
\end{equation}
where $\tilde{h}(f)$ represents the Fourier transform of a time domain signal. To find $\tilde{h}(f)$, we use the phenomenological waveform \textit{PhenomD} \citep{Husa2016, Khan2016}. \textit{PhenomD} is based on fitting analytical templates to numerical relativity waveforms. For a detailed description of its constructions, please see \cite{Husa2016} and \citet{Khan2016}. Here, we focus on how the waveform is determined based on the parameters of the MBHs in our population.

To generate the waveforms, we use the \texttt{gwsnrcalc} Python package from the BOWIE analysis tool \citep{Katz2018}. \texttt{gwsnrcalc} takes as inputs the masses of the MBHs, $M$, $m$; the dimensionless spin of each MBH $a_1$, $a_2$; the redshift of the binary, $z$; and the start and end times of the binary's orbit, in relation to the merger of the binary, $t_\text{st}$ and $t_\text{end}$. 

\begin{figure}
\begin{center}
\includegraphics[scale=0.33]{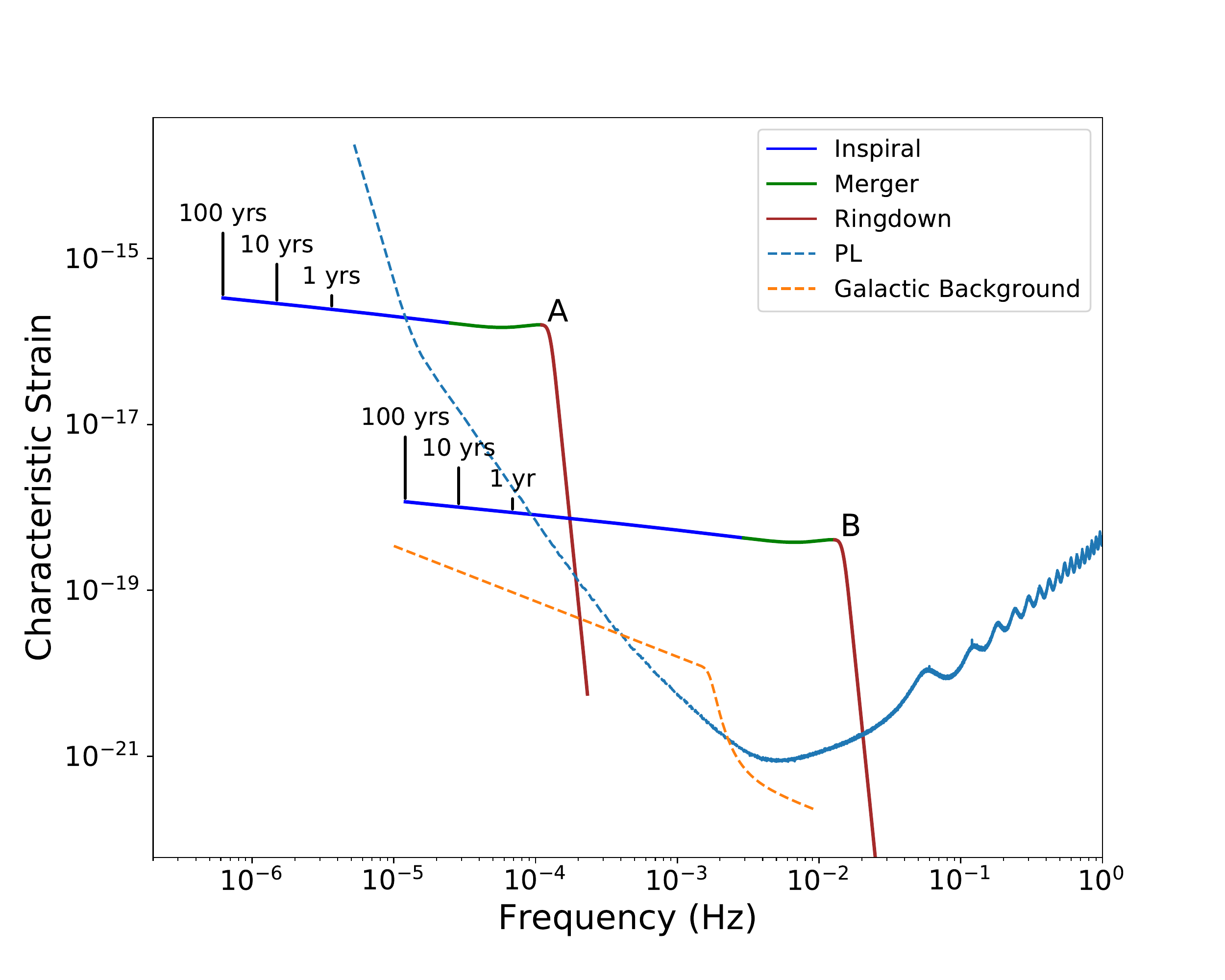}
\caption{Two examples of the characteristic strain, $h_c$, curves are shown here with solid lines. The blue, green, and red portions of the binary signals represent the construction we use for the inspiral, merger, and ringdown, respectively. Both examples show $a=0.8$ and $q=0.2$ for a signal beginning 100 years before merger. To plot these curves, we use $t_\text{st}=100$ yrs and $t_\text{end}=0$ so that we encapsulate 100 years of inspiral as well as the merger and ringdown. The times before merger are labeled above the strain curve for 100, 10, and 1 yrs before merger. Example A shows a binary of $M_T=10^8M_\odot$ and $z=0.75$. Example B shows $M_T=5\times10^5M_\odot$ and $z=2$. In addition to binary signals, the sensitivity curve tested in this work (PL) is shown in characteristic strain of the noise, $h_N$ \citep{LISAMissionProposal}. Additionally, the Galactic background noise we use is shown with a dashed orange line.} \label{fig:sensitivity_curves_and_signals}
\end{center}
\end{figure}

The dimensionless spin of each MBH is $a_i=J_i/m_i^2$, where $J$ is the magnitude of the spin angular momentum. $a$ ranges from -1.0 (anti-aligned to the orbital angular momentum) to 1.0 (aligned to the orbital angular momentum). For convenient use of \textit{PhenomD}, we treat the spins as aligned. Measurements of MBH spins have shown spins near maximal  \citep{Miller2007, Reynolds2013}. For this reason, we choose to model spins of  $a_1=a_2=a=0.8$. As the spin magnitude is raised, the waveform will gain more signal. For near-equal mass systems, which represent a majority of systems in our catalog, the difference in the spin does not change the signal significantly. For systems of mass ratio farther from unity, the spin can have a significant impact on their detectability because the signal peak can increase by an order of magnitude from the spin-down ($a=-1$) to the spin-up case ($a=1$). Therefore, applying this spin configuration ($a=0.8$) represents the optimistic case for these systems. The choice to use the same spin for both MBHs is made because \textit{PhenomD} was calibrated in mostly equal-spin configurations. Within its calibration range, \textit{PhenomD} performs accurately matching waveforms to better than $\sim1\%$ error. Outside of its calibration range, it produces physically reasonable results, indicating it can be useful for basic studies \citep{Khan2016}. See Figure \ref{fig:sensitivity_curves_and_signals} for examples of characteristic strain curves.

\subsubsection{Start and End Times}\label{sec:startendtimes}

\begin{figure}
\begin{center}
\includegraphics[scale=0.58]{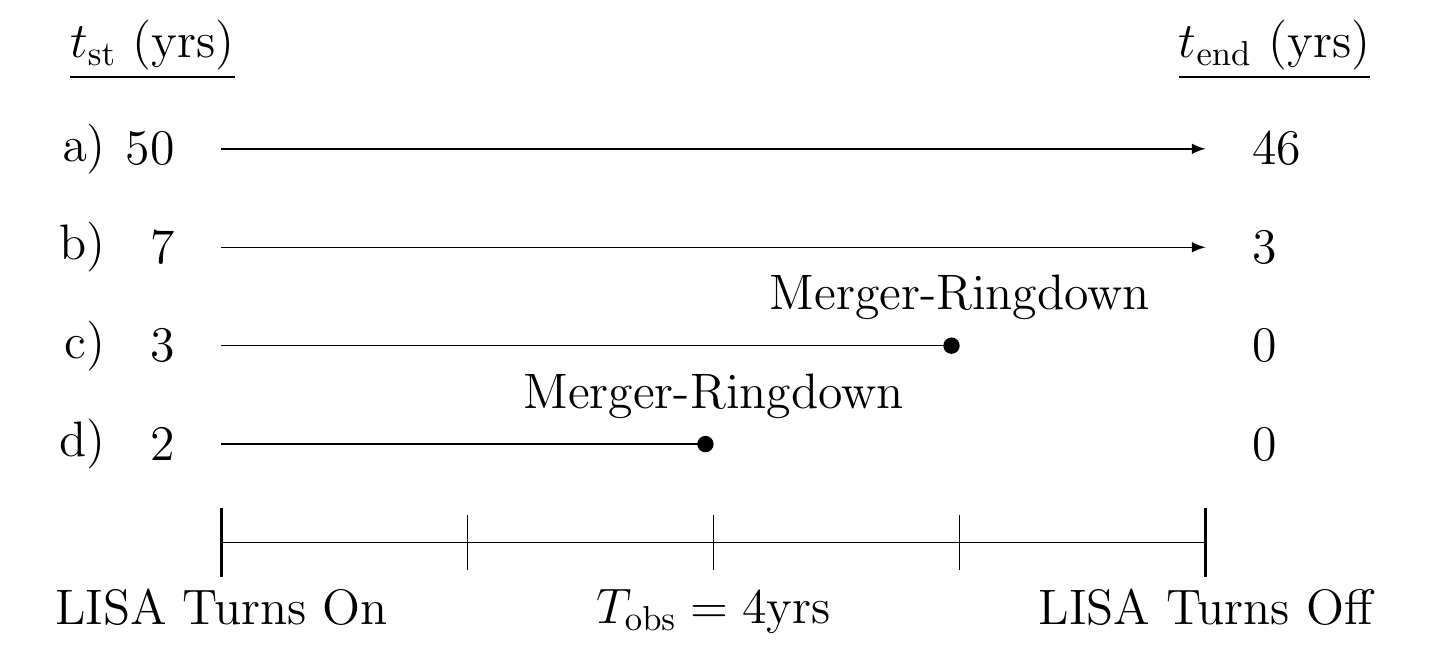}
\caption{Our construction for $t_\text{st}$ and $t_\text{end}$, given in section \ref{sec:startendtimes}, is illuminated with this diagram. We show two binaries (a,b) for which LISA will only measure the inspiral signal because the binary remains far from merger when LISA is turned off. For these binaries the difference between  $t_\text{st}$ and $t_\text{end}$ will be $T_\text{obs}$. With $T_\text{obs}=4$yrs, we will accumulate signal for 4 years as the binary inspirals toward each other. Cases (c) and (d) represent binaries that merge during LISA observation. For these binaries, $t_\text{end}$ is always zero; $t_\text{st}$ determines the duration of time that LISA observes these sources.  }\label{fig:time_to_merger_explanation}
\end{center}
\end{figure}

The start times, $t_\text{st}$, and end times, $t_\text{end}$, both represent the time until merger for a specific binary at which LISA begins and ends its observation of the binary's signal. These times are effectively a map to the frequency bounds of the \textit{PhenomD} waveform model. These times will be on the order of years. Since the merger and ringdown timescale is on the order of minutes to hours for MBHs, we do not include this timescale in $t_\text{st}$ and $t_\text{end}$. The main reason for this construction is it allows us to refrain from assuming a specific observation time for each binary: it allows us to test, within our Monte Carlo sample, binaries that merge at early and late times in the LISA observation window, as well as binaries that merge after the LISA observation window. Early merging binaries will have less time where their inspiral signal can be observed, compared to later mergers. Similarly, LISA will not be able to detect the merger or ringdown for binaries that merge after the LISA observing window, leading to inspiral-only signals (if the inspiral is detectable over the observation time). Figure \ref{fig:time_to_merger_explanation} displays a diagram showing how $t_\text{st}$, $t_\text{end}$, and $T_\text{obs}$ are related for various sources. Please see \mbox{Section \ref{sec:lisaconfig}} in the Appendix for our analysis related to inspiral-only signals.

%Sources that merge within the LISA observation window will have $t_\text{st}<T_\text{obs}$ and $t_\text{end}=0$. This indicates the waveform will exhibit the inspiral, merger, and ringdown. For example, with $T_\text{obs}=5$yrs, if $t_\text{st}<5$yrs, the merger signal will be measured during LISA observation. If $t_\text{st}=5$yrs, the merger signal will occur at exactly the point where the detector is turned off. If a source does not merge during the observation window, $t_\text{st}$ will be its time before merger at the start of the observation window and $t_\text{end}=t_\text{st}-T_\text{obs}$. In this case, the waveform will only come from the inspiral stage. An example would be a source with $t_\text{st}=7$yrs. In this case, $t_\text{end}=2$yrs. This construction allows us to analyze varying values for inspiral duration as well as sources that never reach the merger stage. This gives us a sense of how many sources are observed evolving during inspiral at lower frequencies near the low-frequency band edge of the LISA sensitivity curve. The shape of this low-frequency edge will greatly influence our ability to measure these inspiralling binaries prior to their merger \citep{Katz2018}. Figure \ref{fig:time_to_merger_explanation} displays a diagram showing how $t_\text{st}$, $t_\text{end}$, and $T_\text{obs}$ are related for various sources.

As a point of reference, the merger frequency separating inspiral from merger, is given by
\begin{equation}\label{eq:fmrg}
	f_\text{mrg} = \frac{c^3}{G}\frac{1}{6^{3/2}\pi M_\text{T}\left(1+z\right)},
\end{equation}
where $M_\text{T}=M+m$ is the total mass of the binary in its source frame. The $(1+z)$ term redshifts this mass to the detector frame.

The frequency at a time before merger to 1PN order is given by \citep{Blanchet2014},
\begin{equation}
	f(t) = \frac{c^3}{G}\frac{1}{8\pi M_T(1+z)\tau^{3/8}}\left(1+\left(\frac{11}{32}\eta+\frac{743}{2688}\right)\tau^{-1/4}\right),
\end{equation}
where,
\begin{equation}
	\tau=\frac{c^3}{G}\frac{\eta t}{5M_T(1+z)}.
\end{equation}
$t$ is given in the detector frame. The start frequency of the waveform, $f_\text{st}$, is, therefore, $f(t_\text{st})$. For signals that exhibit inspiral, merger, and ringdown, $t_\text{end}$ is  zero. In this case the end frequency, $f_\text{end}$, is the highest frequency used in the \textit{PhenomD} model, representing the end of the ringdown, given by \citep{Khan2016},
\begin{equation}
	f_\text{end}=\frac{G}{c^3}\frac{0.2}{M_T(1+z)}.
\end{equation}
If $t_\text{end}$ is not zero, indicating that the source is only detected in the inspiral stage, $f_\text{end}$ is $f(t_\text{end})$.

\subsubsection{LISA Sensitivity}\label{sec:LISAsense}

The LISA sensitivity configuration used is from the LISA Mission Proposal \citep{LISAMissionProposal}. We refer to this sensitivity as ``PL'' in \citet{Katz2018}. This sensitivity is based on a 3-arm triangular configuration with 2.5 million km armlengths, 30 cm diameter telescopes, and 2 W end-of-life laser power. In \mbox{Section \ref{sec:lisaconfig}} of the Appendix, we perform an analysis comparing PL to an older iteration of the LISA configuration \citep{Larson2000} to show how the LISA configuration changes will affect various aspects of LISA MBH analysis.

We show our sensitivity curve in \mbox{Figure \ref{fig:sensitivity_curves_and_signals}} in terms of the sky-averaged characteristic strain, $h_N$. The sky-averaging factor is 3/20 \citep{Robson2019}. Sensitivity curves are generally presented in terms of the power spectral density (PSD) of the noise, $S_N$. To convert from $S_N$ to $h_N$, we use $h_N=\sqrt{fS_N}$ \citep{Moore2015}.

We also include the effect of the Galactic background noise in addition to the instrumental noise. We use the analytical approximation of \citet{Hiscock2000} to the Galactic background noise suggested in \citet{HillsBender1997}. This is shown in Figure \ref{fig:sensitivity_curves_and_signals}. Compared to recent predictions from \citet{Robson2017}, this is a conservative estimate of this noise contribution. The contribution of this background can be decreased with proper global fitting methods and a longer observation window \citep{Robson2017}.

\subsubsection{Signal-to-Noise Ratio}\label{sec:SNR}

We use the SNR to determine the detectability of the sources in our catalog. The SNR is estimated by integrating the ratio of the signal to the noise in the frequency domain. The sky, orientation, and polarization averaged SNR is given by \citep{Robson2019},
\begin{equation}\label{eq:SNR}
	\langle\rho^2\rangle = \frac{16}{5} \int_0^\infty \frac{h_c^2}{h_N^2}\frac{1}{f}df. 
\end{equation}
The SNR is then multiplied by a factor of $\sqrt{2}$ because we consider a 2 channel interferometer. 

An additional question we analyze is how many sources exist in our models with a high enough SNR to perform black hole spectroscopy. To do this, we use the General Likelihood Ratio Test (GLRT) formalism suggested in \citet{Berti2016}. Using the GLRT, the SNR of the $l=m=2$ ringdown mode is used as a proxy for determining the detectability of the $l=m=3$ or $l=m=4$ modes. Sources can be spectroscopically measured if $\rho_{\text{l=m=2}}>\rho_{\text{GLRT}}\equiv\text{min}\left(\rho_{\text{GLRT}}^{2,3}, \rho_{\text{GLRT}}^{2,4}\right)$, where $\rho_{\text{GLRT}}$ for each mode is is given by \citep{Berti2016},
\begin{align}\label{eq:rhoGLRT}
	\rho_{\text{GLRT}}^{2,3}&=17.687+\frac{15.4597}{q-1} - \frac{1.65242}{q}, \\
    \rho_{\text{GLRT}}^{2,4}&=37.9181+\frac{83.5778}{q} + \frac{44.1125}{q^2} + \frac{50.1316}{q^3}.
\end{align}

\subsection{Monte Carlo Analysis}\label{sec:montecarlo}

We use a Monte Carlo analysis technique based on Poisson statistics to characterize the range of possibilities resulting from the Illustris output. We do this for multiple reasons. The primary reason is the Illustris output is one iteration of the evolution of a fractional volume within the Universe. We want to understand how the detection rate and source characteristics will vary with the Illustris output as the statistical backdrop. Additionally, we wanted to create a catalog generator for LISA MBH binary signals, which requires a Monte Carlo draw of a new sample each time. As we will discuss in section \ref{sec:poisson}, the Monte Carlo sampling allows us to refrain from assuming an observable duration of the waveform for each of the binaries. Most detection rate predictions assume an observable time for each MBH of 1 year. This does not account for binaries that will merge before one year of LISA observation. It also does not include the longer measurement of an inspiral signal if the binary signal is observable at times longer than 1 year before merger. In other words, our method provides a more realistic basis for assessing binary detectability during the LISA mission.

\subsubsection{Merger Rate Prediction}\label{sec:mergerrate}

The first parameter in the Monte Carlo sampling process is the coalescence rate of MBH binaries in the Illustris simulation. We calculate this parameter one time for each evolutionary prescription. At this stage of the sampling process, we are not considering detectability; we consider any coalescence that occurs prior to $z=0$. For the ND and ND-6 models, all binaries in our sample coalesce before $z=0$ because these two models assume no inspiral time for all MBH binaries. For the binary inspiral models of DA17 and K17, we find that 84\% and 66\% of binaries coalesced before $z=0$, respectively. See section \ref{sec:results-lifetimes} and Figure \ref{fig:coalescence_rates} for more information on how the inspiral models affected the population of binaries coalescing before $z=0$. We determine the number of coalescences, $N$, in a given redshift interval $z+\Delta z$. We then compute the number of coalescence events across redshift intervals per comoving volume element,
\begin{equation}\label{eq:dn_dzdVc}
	\frac{d^2 n(z)}{dzdV_c} \approx \frac{N(z)}{\Delta z V_c},
\end{equation}
where $V_c$, the term on the right hand side, is the comoving volume of the Illustris simulation, (106.5 Mpc)$^3$. We can then calculate the number of coalescences per observing time interval given by, 
\begin{equation}\label{eq:coalrate}
	\frac{dN_{\text{coal}}}{dt_{\text{obs}}} = \int_0^\infty \frac{d^2\bar{n}(z)}{dzdV_c} \frac{dz}{dt}\frac{dV_c}{dz}\frac{dz}{1+z},
\end{equation}
where the $1+z$ redshifts the infinitesimal time element in $dz/dt$ to the observer frame time interval. When we refer to the ``integral rate calculation,'' we are referring to \mbox{Equation \ref{eq:coalrate}}. For our Monte Carlo catalogs, this quantity becomes our input into our Poisson rate calculator.

\subsubsection{Poisson Sampling}\label{sec:poisson}

The two parameters needed to perform the desired Poisson sampling is rate of coalescences determined from \mbox{Equation \ref{eq:coalrate}} and the duration for which we want to draw potential sources, $t_{\text{dur}}$. If we only wanted to draw sources for the observation window, we would set $t_{\text{dur}}=T_{\text{obs}}$. However, this would only focus on sources coalescing within the observation window. We also want to test for inspiraling sources that would coalesce some time after the LISA observing window. Therefore, we choose $t_{\text{dur}}>T_{\text{obs}}$. We tested a variety of values for $t_{\text{dur}}$. We found $t_{\text{dur}}=10^2$ yrs encompassed all of the observable systems in our catalogs, while maintaining computational efficiency.

Our final Poisson parameter, $\lambda$, is given by,
\begin{equation}
	\lambda = \frac{dN_{\text{coal}}}{dt_{\text{obs}}}t_{\text{dur}}.
\end{equation}
In other words, this is the expected number of coalescence events over $10^2$ years. For each catalog, we draw the number of sources occurring within our 100 year window from this Poisson distribution.

\subsubsection{Event Times}\label{sec:eventtimes}

When we have the number of sources drawn, we assign each a random coalescence time, $t_{\text{ev}}$, between zero and $10^2$ yrs. At the timescales we are considering ($\sim10^2$ yrs), the distribution of sources over time will not be affected by the evolution of the Universe. $t_{\text{ev}}=0$ indicates an event occurring at the moment LISA begins observations. $t_{\text{ev}}=T_{\text{obs}}$ represents an event occurring at the moment the LISA observation window ends.

When considering the waveform described in \makebox{section \ref{sec:charstrain}}, we use $t_{\text{ev}}$ and $T_{\text{obs}}$ to determine the start ($t_{\text{st}}$) and end ($t_{\text{end}}$) times related to waveform creation. $t_{\text{st}}=t_{\text{ev}}$ because the event time represents the time before merger at the start of LISA observation. Therefore, $t_{\text{end}}=t_{\text{ev}} - T_\text{obs}$ if $t_\text{ev}>T_\text{obs}$. If $t_\text{ev}\leq T_\text{obs}$, $t_{\text{end}}=0$ (see Figure \ref{fig:time_to_merger_explanation}).

\subsubsection{Resampling Binary Parameters}\label{sec:resampling}

After sampling the number of binaries and the event times for each event, we need to sample binary parameters of $M$, $m$, and $z$. To do this, we use kernel density estimation methods. However, there is a key distinction that needs to be made for sampling these binary parameters: we must incorporate the volume and time redshifting factors implicit in the expansion of the universe as weights, $w$, in the density estimation. If you assume an infinitesimal redshift bin width for \mbox{Equation \ref{eq:coalrate}} ($\Delta z\rightarrow dz$), this weighting factor as a function of redshift is given by,
\begin{equation}\label{eq:weights}
	W(z)=\frac{dz}{dt}(z)\frac{dV_c}{dz}(z)\frac{1}{1+z}.
\end{equation}
 The weight applied to the $i$th binary is then \mbox{$w_i = W_i/\sum_{i=1}^NW_i$}. We also include the covariance across these parameters in the KDE, so as to sample the population accurately.

\section{Results}\label{sec:results}
\subsection{Binary Lifetimes}\label{sec:results-lifetimes}

\begin{figure*}
\begin{center}
\includegraphics[scale=0.6]{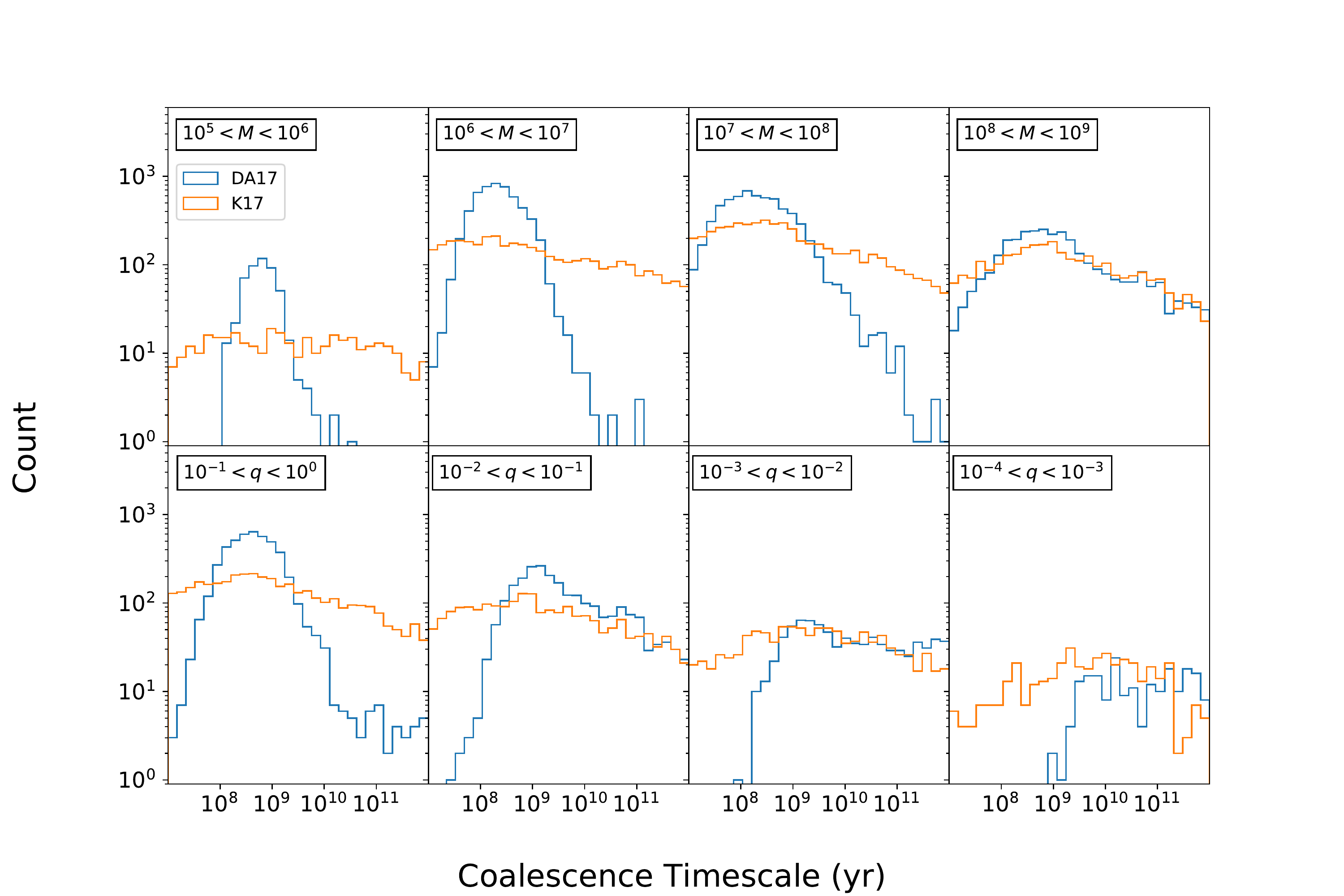}
\end{center}
\caption{Coalescence timescales are shown for the DA17 and K17 models in blue and orange, respectively. The top row shows binaries grouped by decades in total mass, $M_T$. The bottom row shows binaries grouped by decades in mass ratio, $q$.}
\label{fig:lifetimes}
\end{figure*}

\begin{figure*}
\begin{center}
\includegraphics[scale=0.4]{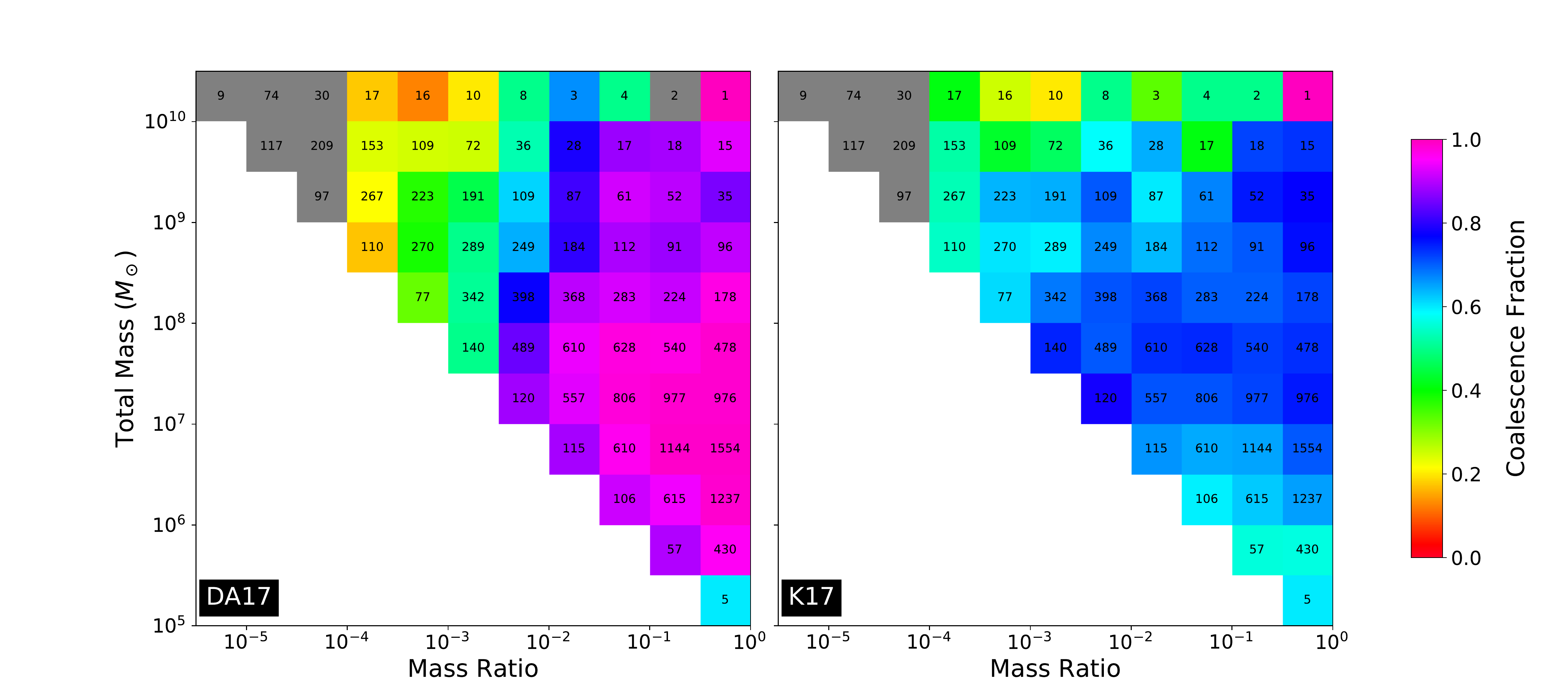}
\end{center}
\caption{Coalescence fractions are compared for the DA17 (left) and K17 (right) models. These fractions are binned in total mass and mass ratio. The number in each bin represents the total number of binaries residing in that bin. Therefore, this is the same for both models. The color represents the coalescence fraction based on the color bar on the right. None of the binaries shown in grey coalesce before $z=0$. The white space represents binaries not analyzed here due to the Illustris resolution limit.}
\label{fig:coalescence_rates}
\end{figure*}

We first test and compare our evolutionary prescriptions to understand how the initial population of binaries will change when evolved to coalescence with different sub-grid models. In the ND and ND-6 models, all binaries are considered to be coalesced prior to $z=0$. By modeling the sub-grid physics as in the DA17 and K17 models, some binaries will no longer merge before $z=0$ and will, therefore, deflate the merger rate. Additionally, the mass distributions of the coalesced binaries may change because the prescriptions have different dependencies on the masses. Figure \ref{fig:lifetimes} shows the evolutionary timescales calculated for all binaries binned by total mass and mass ratio. The DA17 model generally results in a more peaked distribution, while K17 shows a flatter profile across all plots. However, at higher total masses, the two prescriptions become very similar in their predictions. Figure \ref{fig:coalescence_rates} shows the effect of these evolutionary timescales on the coalescence fractions of our population. This figure illustrates the global differences, in terms of binary parameters, between the two prescriptions. DA17 favors near-equal mass and low total mass systems, while K17 favors near-equal mass systems with total masses towards the higher end. With the K17 model, larger masses are favored because they are embedded in higher density, more centrally concentrated stellar cores. On the other hand, the specific dynamical friction prescription used in the DA17 model \mbox{(see Equations \ref{eq:tdotbare}, \ref{eq:tdotgx}, and \ref{eq:tdotmin})} causes high-mass systems to exist for longer times in the dynamical friction stage between the influence radius of the larger MBH \mbox{(Equation \ref{eq:rinfl})} and the hardening radius \mbox{(Equation \ref{eq:a_h})} (please see the original papers for more details). It is also clear the overall coalescence fractions are higher with DA17 than with K17. The overall coalescence fraction for DA17 was 84\%. With the K17 model, only 66\% of all binaries coalesced before $z=0$.

\subsection{Rate Predictions}\label{sec:results-mergerrate}

\begin{figure*}
\begin{center}
\includegraphics[scale=0.45]{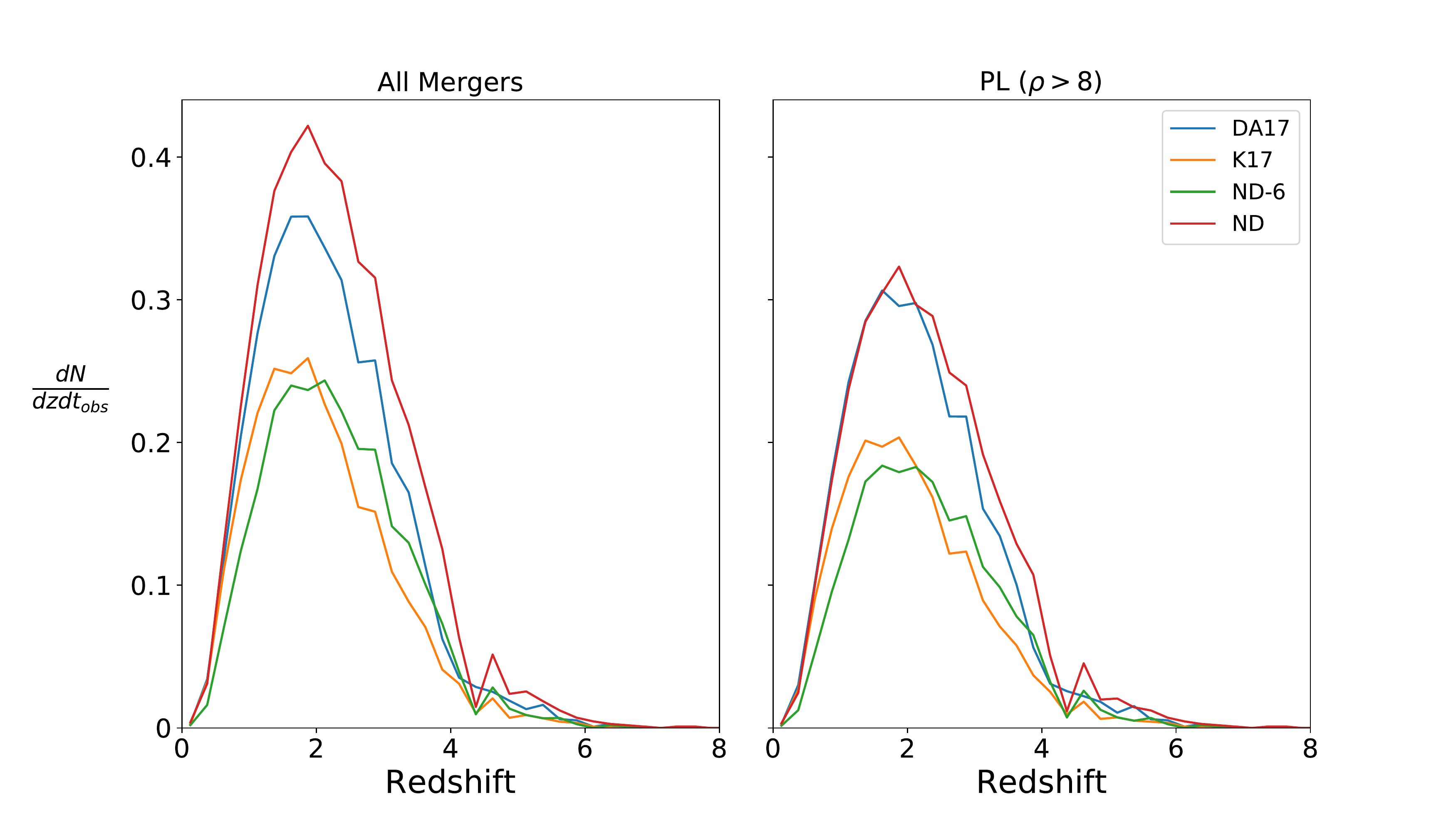}
\end{center}
\caption{Merger rates per year per unit redshift are shown above. ND (no delays), ND-6 (subset of ND model with $m_1,m_2\geq10^6M_\odot$), K17 \citep{PTA-illustris, Kelley2017b}, and DA17 \citep{Dosopoulou2017} models are shown in red, green, orange, and blue, respectively. The left plot shows all mergers from the simulation with each prescription. The right shows the detection rate per year per redshift for each model assuming an SNR cut of $\rho=8$.}
\label{fig:detection_results}
\end{figure*}

We calculate merger rates with two methods: integrating the redshift distributions with Equation \ref{eq:coalrate} and Monte Carlo sampling (section \ref{sec:montecarlo}). The results are similar as expected; however, the Monte Carlo aspect allowed for more freedom in terms of not setting specific values for observation duration and starting times before merger. It also ensures we are examining different realizations of the merging MBH population. Table \ref{tb:integral_rates} shows integral merger rate results for each prescription. It also shows detection rates for each stage of binary black hole coalescence as well as the rate of sources where spectroscopic measurements (Equation \ref{eq:rhoGLRT}) are possible. Similar results for the Monte Carlo method are shown in \mbox{Table \ref{tb:monte_carlo_rates}}. The redshift distributions used in these calculations are shown in Figure \ref{fig:detection_results}. For the main integral merger rate and detection rate results, the rate is quoted as per year. For the following integral rate calculations, we estimate our standard deviation in our predictions to be less than 0.01 (based on our chosen redshift bin width), according to Equation 17 in \citet{Salcido2016}. %$T_\text{obs}$ was assumed to be 1 year. 

First, comparing ND with ND-6 (``ND-6'' represents the subset of binaries in ``ND'' which have each constituent mass above $10^6M_\odot$), we see the advanced extraction was important for LISA-related analysis since the ND merger rate was almost two times the merger rate of ND-6. This holds true for the detection rates as well. An interesting statistic here is the inspiral detection rate, 0.23 yr$^{-1}$ with ND-6 versus 0.45 yr$^{-1}$ with ND, as this demonstrates the importance of the lower-mass systems retained by our advanced extraction technique.

Comparing all four models, we see some interesting results. The hierarchy predicted in section \ref{sec:lifetimes} is apparent. The DA17 rates resemble much more strongly the ND rates compared to K17 due to the inability of K17 to coalesce the large number of low mass systems. Interestingly, the K17 rates strongly resemble the ND-6 rates, indicating the loss of the low mass systems, as well as the general loss in coalescing systems due to adding sub-grid modelling, caused the rate to decrease to similar levels as without the low mass systems entirely. These aspects can be seen clearly in \mbox{Figure \ref{fig:detection_results}}. After calculating the SNR and making our cut at $\rho=8$, we see that the DA17 curve tracks the ND curve, while the K17 curve mirrors the ND-6 curve.

%When comparing the two LISA configurations, we see the CLLF configuration attains a rate that is roughly 10\% larger than PL. Every binary that is detected will have a detectable merger-ringdown, which can be seen in the identical rates between full-signal rates and merger-ringdown rates for both PL and CLLF.

\begin{figure*}
\begin{center}
\includegraphics[scale=0.45]{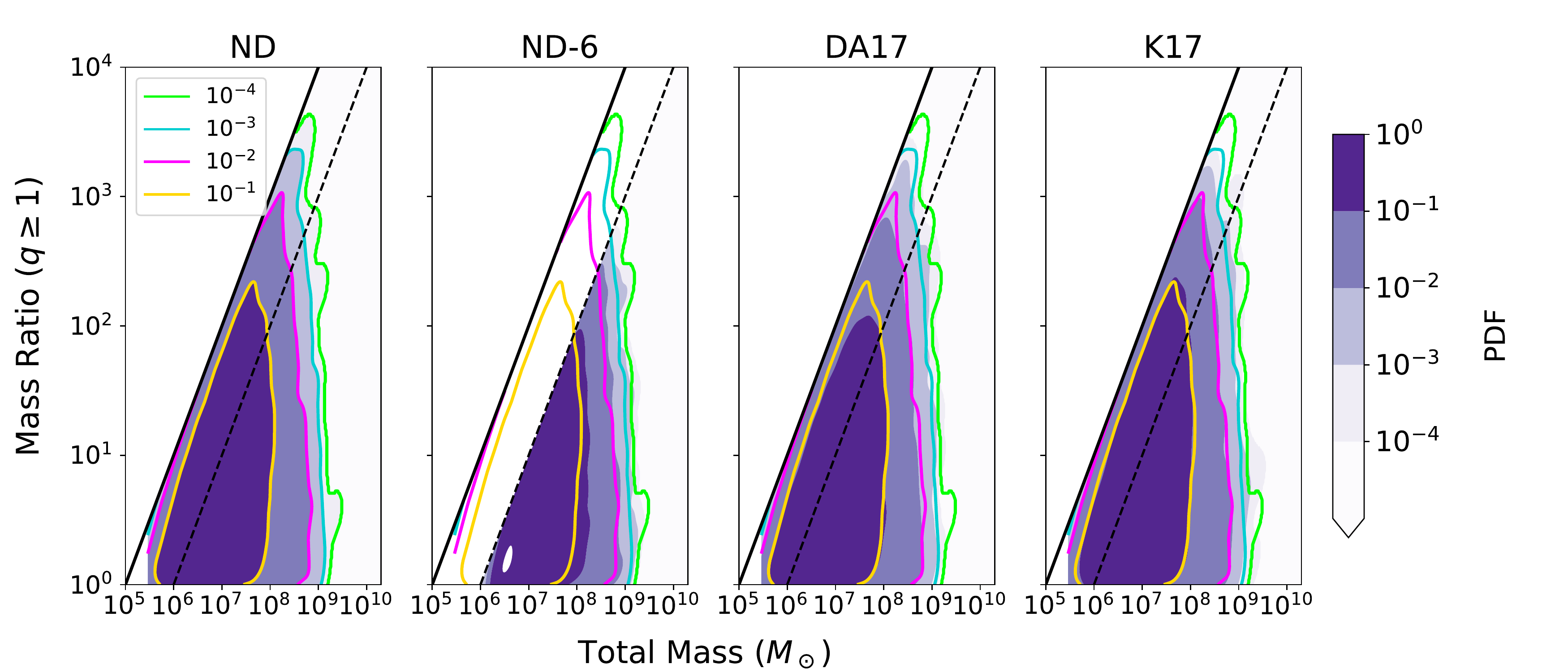}
\end{center}
\caption{Probability density functions (PDF) are shown for mass ratios and total masses of observed binaries ($\rho\geq8$) from our 10000 Monte Carlo catalogs. The colored, filled contours show the PDF for the model given in the title of each plot. The colored line contours represent the PDF of the ND model, which we overplot in each panel. The left plot shows the ND (``no delays''). In the center-left, we compare ND-6 with the ND model. The ND-6 model is a subset of the ND model with $m_1,m_2\geq10^6M_\odot$.  We then show the DA17 and K17 models as the center-right and right plots, respectively. When comparing DA17 and K17 to the ND model, there is minimal discernible difference between the three. Empirically, their means and higher order moments in the total mass and mass ratio are within a small percentage of one another. Slightly higher mass ratios are more prevalent in the ND and K17 models compared to the DA17 model. The solid and dashed black lines show the effect of a mass cutoff at $10^5M_\odot$ and $10^6M_\odot$, respectively.}
\label{fig:mt_vs_mr}
\end{figure*}

When analyzing the mass distributions, we find all models produce similar results. Mass distributions from our Monte Carlo analysis can be seen in Figure \ref{fig:mt_vs_mr}. The black solid line shows the limit imposed by the Illustris simulation seed mass at $\sim10^5M_\odot$. Binaries with total mass and mass ratio values above this line cannot exist. Similarly, the black dashed line shows the limit imposed by the $10^6M_\odot$ cut. This once again highlights the effect of the more advanced extraction. ND, DA17, and K17 exhibit roughly the same structure. Specifically, their mean values and higher order moments about the mean values are within a small percentage of each other; however, DA17 has a slightly smaller kurtosis in the mass ratio. This general similarity does indicate the K17 model has a relatively flat effect across the parameter space, where sources are detectable by LISA, suppressing each mass and mass ratio regime in an equivalent manner.

\begin{table}
\centering
 \begin{tabular}{||c c c c c c||} 
 \hline
 Prescription & Merger Rate & All & Ins & MR & BHS \\ [0.5ex] 
 \hline
ND & 0.98 & 0.75 & 0.45 & 0.77 & 0.74  \\
\textit{ND-6} & \textit{0.57} & \textit{0.44} & \textit{0.23} & \textit{0.45} & \textit{0.43} \\
DA17 & 0.80 & 0.70 & 0.42 & 0.70 & 0.67 \\
K17 & 0.55 & 0.44 & 0.28 & 0.45 & 0.43 \\
 \hline
\end{tabular}
\caption{Merger and detection rate calculations per year are shown using the integral calculation (Equation \ref{eq:coalrate}). The evolution prescriptions are listed in the first column. The top row shows the ``no delays'' model: ND. The second row shows the ND-6 model, displayed in italics because it represents a subset of the ND model with $m_1,m_2\geq10^6M_\odot$. The final two rows show the DA17 \citep{Dosopoulou2017} and K17 \citep{PTA-illustris, Kelley2017b} models, respectively. The merger rate gives the rate of coalescences without considering LISA detectability. The remaining columns show the detection rate ($\rho>8$). Additionally, detection rates are separated into signal types: All, Ins, MR, and BHS. ``All'' indicates reaching detection threshold using the entire signal. ``Ins'' and ``MR'' represent detection rates of inspiral signals and signals from the merger and ringdown, respectively. These categories are not independent: a single binary can add to the rate in both categories. ``BHS'' is the detection rate of MBH binaries where black hole spectroscopy is possible (see Equation \ref{eq:rhoGLRT}). We estimate our errors in these predictions to all fall below 0.01 according to equation 17 in \citet{Salcido2016}.}
\label{tb:integral_rates}
\end{table}

\section{Discussion}\label{sec:discussion}

\begin{table*}
\centering
 \begin{tabular}{||c c c c c c c c c c||} 
 \hline
 \hline
 Reference & Base Population & MBHB Evolution Prescription & Merger Rate (yr$^{-1}$) \\
%  \hline
%  \citet{Sesana2004} & SAM & DF,LC & $\sim20$ \\
 \hline
  \citet{Arun2009} & SAM & None & $\sim22$ \\
 \hline
 \citet{Sesana2011} & SAM & None & $\sim25$ \\
 \hline
\citet{Klein2016} & SAM & DF,LC,VD,GW,Tri & $\sim$8 \\
 \hline
\citet{Berti2016}$^1$ & SAM & DF,LC,VD,GW,Tri & $\sim$8 \\
 \hline
\citet{Salcido2016} & Hydrodynamic & Constant$^2$ & $\sim$2 \\
 \hline
 \citet{Bonetti2019} & SAM & DF,LC,VD,GW,Tri & $\sim$23 \\
 \hline
 This Paper & Hydrodynamic & DF,LC,VD,GW & $\sim0.5-1$ \\
 \hline
 \hline
\end{tabular}
\caption{A collection of papers quoting rates for the detection of MBH binaries by LISA is shown above. We focus here on papers analyzing the large-seed ($10^4-10^6M_\odot$) formation channels since this is similar to the seeding mechanism in Illustris. The base population tells if the models are based on SAMs or hydrodynamic simulations. Short descriptions of the MBH binary evolutionary prescriptions are also given. ``None'' indicates that no delays between galactic and MBH binary mergers were included. ``Constant'' indicates that a constant delay was used for all binaries. ``DF'' is a dynamical friction prescription. ``LC'' indicates inclusion of a stellar hardening or loss-cone scattering model. A prescription involving torque from a gas disc is indicated with ``VD.'' ``GW'' indicates a gravitational wave driven inspiral. A prescription containing triple MBH systems is expressed with ``Tri.''  \newline $^1$For detectable ringdown signals. For the mass ranges we consider in this paper, ringdown signals will be measurable for all detectable sources. \newline $^2$A constant delay was chosen based on gas-heavy versus gas-poor host galaxies.}
\label{tb:summary_table}
\end{table*}

The merger rates and LISA detection rates of MBH binaries from the Illustris simulation are low compared to the majority of the literature on this subject. Table \ref{tb:summary_table} summarizes predicted rates in the literature, including whether their base population is from a semi-analytical model (SAM) or a hydrodynamic simulation. Additionally, the type of delay prescription employed is mentioned. For this summary, we focused on predictions related to high-mass seeds since this seeding prescription better matches our setup in this paper. Even with our ND model, which represents the rates directly from the simulation without any sub-grid modeling, our predicted rate is low. LISA will be sensitive to binaries $\leq10^5M_\odot$. Therefore, any study that cannot resolve the galaxies and halos where these low masses evolve \citep[see section D in ][]{Klein2016} will under-predict the total detection rate. Examining Figure \ref{fig:mt_vs_mr}, we see LISA primarily observes binaries of $1\leq q \leq10^2$. Therefore, if we consider our population to be complete when we can resolve all constituent MBHs at mass ratios up to $10^2$, then our study is only complete above $10^7M_\odot$ leading to a further underestimation of the overall detection rate. Another way to consider this effect is by understanding how the seeding prescription will affect the base population. Without seeds below $10^5M_\odot$, the detection rate for masses above $10^5M_\odot$ will also be deflated: seeds below this mass may grow to a mass above $10^5M_\odot$ before undergoing a merger. This aspect significantly affects our merger rate prediction as it removes MBHs from our population at the most common and detectable masses in our sample.

The literature references for SAMs shown in Table \ref{tb:summary_table} employ seeds of $\sim10^4M_\odot$. Within the subset of these SAMs predictions, the higher rate predictions occur with no binary inspiral model employed (similar to our ND model). However, early SAMs predictions did analyze various delay prescriptions, but for smaller seeds ($\sim150M_\odot$) \citep[e.g.][]{Volonteri2003, Sesana2004, Sesana2005}. Historically, SAMs have predicted rates about an order of magnitude larger than our findings; however, our results are within an order of magnitude of those predicted in \citet{Salcido2016} for the EAGLE simulations \citep{EAGLE-release}, indicating similarity between two hydrodynamic-based populations. While hydrodynamic simulations strive to produce populations from simple physics, they are restricted to a much more limited parameter space than SAMs (discussed further below). This means that rate predictions from SAMs and hydrodynamic-based models fundamentally differ.

One clear difference is the ability of SAMs to explore an arbitrarily large range of masses, while hydrodynamic simulations are resolution limited, typically to above $\sim10^5M_\odot$. If we consider the increased prevalence of MBHs as we move towards smaller masses, we would expect that models with access to binaries below $10^5M_\odot$ would greatly inflate the merger rate in a similar fashion to what we have seen when including the lower masses from Illustris through our advanced extraction. Along the same lines, Illustris, and simulations like it, do not access the scales needed to examine populations of dwarf galaxies. As evidence mounts that dwarf galaxies house MBHs in their center in both observations \citep{Reines2013, Moran2014, Satyapal2014, Lemons2015, Sartori2015, Pardo2016, Nguyen2018, Nguyen2019} and simulations \citep{Volonteri2008, vanWassenhove2010, Bellovary2019}, we must improve models since these dwarf galaxy sources are enitrely missing from our analysis. Due to the prevalence of dwarf galaxies as well as the general understanding that dwarf galaxies consistently merge into larger ``host'' galaxies over time, missing the dwarf galaxy MBHs could deflate our rate calculations significantly. This is especially true because these dwarf galaxies will house smaller MBHs that will produce strong signals in the LISA frequency band. Similarly, a number of potential systems may still be missed as we approach the resolution limit of $\sim10^5M_\odot$. 

Similar to this issue with dwarf galaxies, the MBH seeding mechanism plays a large role in rate calculations. The seeding mechanism in Illustris (and EAGLE) is ad-hoc. The seeding model chosen for Illustris produces seeds at later times than those seen in other simulations like \citet{Tremmel2018}. Additionally, the Illustris seeding prescriptions will produce seeds at much later times when compared to SAMs. Later seeding means a smaller volume accessible to LISA observations. Therefore, merger rates from SAMs may also be intrinsically higher at higher redshits. 

The sub-grid models also affect our rates given that they cause coalescences to occur after $z=0$. The particular model chosen has the potential to deflate the rate significantly. As previously mentioned, SAM predictions decreased as more ``delay'' models were incorporated. In \citet{Tremmel2018}, their simulation works to numerically resolve smaller scales in order to track MBH pair formation. They show that the expected number of close MBH binary pairs is lower than expected, which would lead to a depletion in the LISA rate. However, as seen in \citet{Bonetti2019}, including triple interactions can remedy this issue quite significantly, producing rates similar to those found when not considering a delay prescription ($\sim23$). Additionally, including prescriptions for gas-driven migration can also increase our rates with our DA17 model (it is included in the K17 model). Without the inclusion of gas-driven migration or triple interactions, our rate is truly a lower limit. However, it can be seen that inclusion of these prescriptions would not significantly alter our results since the rate from our ND model is only slightly larger than those found with the DA17 and K17 prescriptions.

 For the EAGLE simulations, \citet{Salcido2016} test seeds of $\sim10^5M_\odot$, similar to the seed model in Illustris. Therefore, we believe two reasons other than the seed mass are the main factors in our predicted rate difference. First, \citet{Salcido2016} perform a different extraction analysis, compared to the analysis done for this paper, to deal with numerical issues similar to those suggested in section \ref{sec:mergerpop}. This may inflate the true number of mergers seen in the EAGLE simulations, especially at masses near the seed mass. Second, we use more detailed binary inspiral models than the model used in \citet{Salcido2016}. In their paper, they choose an inspiral time for all binaries as flat values based on if a galaxy merger was gas-rich (0.1 Gyr) or gas-poor (5 Gyr). When we compare these inspiral times with those from our more detailed models (see Figure \ref{fig:lifetimes}), we see that our models will generally predict longer inspiral times, as well as inspiral times that are longer than the age of the Universe. Therefore, the fraction of coalescing binaries before $z=0$ is larger with their choice of inspiral model, which would lead to a higher predicted detection rate. 
 
If we consider our rates in a relative way, even though our predicted rates are low, the comparison of our binary inspiral models can have an impact on LISA MBH science. We have shown the difference between merger models is relatively small, but still varies by a factor of two. Since most detections are from low-mass systems (see Figure \ref{fig:coalescence_rates}), much of this factor of two difference will be lost in the low-mass regime. This means the K17 binary inspiral prescription will lead to more difficulty characterizing the low-mass population, as well as seeding models which will predominantly be constrained by this lower mass regime.

\section{Conclusions}

We have presented new MBH binary LISA rate calculations based on the Illustris cosmological simulations. Our MBH catalog is determined from the Illustris output using a new advanced extraction method allowing us to probe masses down to the simulation seed mass of $\sim10^5M_\odot$. Previous extraction methods made a mass cut requiring $m_1,m_2\geq10^6M_\odot$. By strictly following the interaction between MBH mergers and the MBH host galaxies, we were able to refine the analysis and retain an additional 8,265 mergers in our sample ($\sim50\%$ of mergers analyzed). This doubled the rate of predicted detections for LISA. Binaries containing an MBH of $\sim10^5M_\odot$, especially near equal-mass binaries, are prime targets of the LISA mission due to their time spent evolving in the LISA band, their high SNR potential, and their ability to better probe MBH seeding models.

With this MBH merger catalog from Illustris, we tested four evolutionary prescriptions to understand how binary inspiral models affect our rate predictions, as well as the binary properties of the detected population. Our base model was the ND model, which represented the exact prediction from the Illustris simulation without any delays between the mergers in the simulation (at $\sim$kpc scales) and true coalescence of the MBH pairs. In order to understand the effect of our new extraction method, we also tested a model without delays while requiring $m_1, m_2\geq10^6M_\odot$. We found the detection rate was diminished by a factor of 2 when we included this mass cut.

In addition to our base models without delays, we tested two recent sub-grid models proposed in \citet{Dosopoulou2017} (DA17; section \ref{sec:fanimodel}) and \citet{PTA-illustris,Kelley2017b} (K17; section \ref{sec:kelleymodel}). When comparing the evolution timescales predicted by these two models, we find that DA17 produces a smaller spread in orders of magnitude of this timescale when looking at different masses and mass ratios. K17 produces a relatively flat profile of timescales when looking at all masses and mass ratios in our catalog. When examining coalescence fractions in Figure \ref{fig:coalescence_rates}, we find that DA17 favors lower total masses while K17 favors higher total masses. They both favor more equal mass ratio binaries. Additionally, it was clear from our analysis that DA17 predicts that more binaries from our catalog will merge prior to $z=0$ (84\%) compared to K17 (66\%). 

Due to the ability to coalesce lower mass binaries, the DA17 model resembled the ND model in terms of detection rate at $\sim0.7$ yr$^{-1}$ (integral rate calculation). The K17 model, with its lower overall coalescence fraction and inability to retain the low-mass binaries, led to a detection rate prediction similar to the ND-6 model at $\sim0.4$ yr$^{-1}$. The rates predicted for black hole spectroscopy were similar in magnitude to the overall detection rates. These rates represent lower limits for similar rate predictions (see \mbox{Section \ref{sec:discussion}}).

%With our integral rate calculation, we also tested the ability of LISA to detect sources that do not coalesce within the LISA observation window. We found that CLLF does allow for the measurement of sources in this case, predicting rates of $\sim10^{-2}-10^{-1}$yr$^{-1}$ for binaries within 10 years to merger when LISA is turned on (we assume an observation time of 4 years). This was true for all the models tested with the exception of ND-6. Without any low-mass binaries for measurement earlier in their evolution by LISA, the detection rates for these inspiral-only sources were about an order of magnitude lower. With PL, the detection of these inspiral-only sources is virtually impossible: PL's low-frequency performance does not allow the measurement of these binaries evolving near its low-frequency band edge. 

%Our Monte Carlo results match the predictions laid out in our integral rate predictions. The main takeaway from the Monte Carlo predictions is that $\sim15\%$ of sources detected by CLLF are inspiral-only binaries. Therefore, its low-frequency performance will expand its catalog by approximately this percentage.

We also examined the probability density functions of the total masses and mass ratios of the detected binaries using Monte Carlo generated catalogs. We found that all models with low-mass support (ND, DA17, and K17) produced similar detectable populations in terms of these parameters. This indicates that the surpression of sources by K17 is effectively equivalent across mass regimes, leading to a lower detection rate, while maintaining a similar total mass and mass ratio probability density function to the DA17 and ND models. The ND-6 model cannot match this because it has an entirely inaccessible region where at least one constituent MBH of less than $10^6M_\odot$ is required. Overall, we show that these two detailed models for the evolution of MBH binaries from \citet{Dosopoulou2017} and \citet{PTA-illustris, Kelley2017b} lead to differences in the detection rate and the observable population by LISA.

\section*{Acknowledgements}
We would like to thank Marta Volonteri and the referee for helpful suggestions to improve our paper. M.L.K. acknowledges support from the National Science Foundation under grant DGE-0948017. This research was supported in part through the computational resources and staff contributions provided for the Quest/Grail high performance computing facility at Northwestern University. F.D. acknowledges support from PCTS and Lyman Spitzer Jr fellowships. L.B. acknowledges support from National Science Foundation grant AST-1715413. Astropy, a community-developed core Python package for Astronomy, was used in this research \citep{Astropy}. This paper also employed use of Scipy \citep{scipy}, Numpy \citep{Numpy}, and Matplotlib \citep{Matplotlib}.

%%%%%%%%%%%%%%%%%%%%%%%%%%%%%%%%%%%%%%%%%%%%%%%%%%



%\bibliographystyle{mnras}


\bibliographystyle{mnras}
\bibliography{bibliography_2}

\begin{thebibliography}{}
\makeatletter
\relax
\def\mn@urlcharsother{\let\do\@makeother \do\$\do\&\do\#\do\^\do\_\do\%\do\~}
\def\mn@doi{\begingroup\mn@urlcharsother \@ifnextchar [ {\mn@doi@}
  {\mn@doi@[]}}
\def\mn@doi@[#1]#2{\def\@tempa{#1}\ifx\@tempa\@empty \href
  {http://dx.doi.org/#2} {doi:#2}\else \href {http://dx.doi.org/#2} {#1}\fi
  \endgroup}
\def\mn@eprint#1#2{\mn@eprint@#1:#2::\@nil}
\def\mn@eprint@arXiv#1{\href {http://arxiv.org/abs/#1} {{\tt arXiv:#1}}}
\def\mn@eprint@dblp#1{\href {http://dblp.uni-trier.de/rec/bibtex/#1.xml}
  {dblp:#1}}
\def\mn@eprint@#1:#2:#3:#4\@nil{\def\@tempa {#1}\def\@tempb {#2}\def\@tempc
  {#3}\ifx \@tempc \@empty \let \@tempc \@tempb \let \@tempb \@tempa \fi \ifx
  \@tempb \@empty \def\@tempb {arXiv}\fi \@ifundefined
  {mn@eprint@\@tempb}{\@tempb:\@tempc}{\expandafter \expandafter \csname
  mn@eprint@\@tempb\endcsname \expandafter{\@tempc}}}

\bibitem[\protect\citeauthoryear{Abbott et~al.,}{Abbott
  et~al.}{2017}]{BNSstandardsiren}
Abbott B.~P.,  et~al., 2017, \mn@doi [\nat] {10.1038/nature24471}, 551, 85

\bibitem[\protect\citeauthoryear{Alvarez, Wise  \& Abel}{Alvarez
  et~al.}{2009}]{Alvarez2009}
Alvarez M.~A.,  Wise J.~H.,   Abel T.,  2009, \mn@doi [\apj]
  {10.1088/0004-637x/701/2/l133}, 701, L133

\bibitem[\protect\citeauthoryear{Amaro-Seoane, Eichhorn, Porter  \&
  Spurzem}{Amaro-Seoane et~al.}{2010}]{Amaro-Seoane2010}
Amaro-Seoane P.,  Eichhorn C.,  Porter E.~K.,   Spurzem R.,  2010, \mn@doi
  [\mnras] {10.1111/j.1365-2966.2009.15842.x}, 401, 2268

\bibitem[\protect\citeauthoryear{{Amaro-Seoane} et~al.,}{{Amaro-Seoane}
  et~al.}{2017}]{LISAMissionProposal}
{Amaro-Seoane} P.,  et~al., 2017, arXiv e-prints, \href
  {https://ui.adsabs.harvard.edu/abs/2017arXiv170200786A} {p. arXiv:1702.00786}

\bibitem[\protect\citeauthoryear{{Ardaneh}, {Luo}, {Shlosman}, {Nagamine},
  {Wise}  \& {Begelman}}{{Ardaneh} et~al.}{2018}]{Ardaneh2018}
{Ardaneh} K.,  {Luo} Y.,  {Shlosman} I.,  {Nagamine} K.,  {Wise} J.~H.,
  {Begelman} M.~C.,  2018, \mn@doi [\mnras] {10.1093/mnras/sty1657}, \href
  {https://ui.adsabs.harvard.edu/abs/2018MNRAS.479.2277A} {479, 2277}

\bibitem[\protect\citeauthoryear{Arun et~al.,}{Arun et~al.}{2009}]{Arun2009}
Arun K.~G.,  et~al., 2009, \mn@doi [Classical and Quantum Gravity]
  {10.1088/0264-9381/26/9/094027}, 26, 94027

\bibitem[\protect\citeauthoryear{{Astropy Collaboration} et~al.,}{{Astropy
  Collaboration} et~al.}{2013}]{Astropy}
{Astropy Collaboration} et~al., 2013, \mn@doi [\aap]
  {10.1051/0004-6361/201322068}, 558, A33

\bibitem[\protect\citeauthoryear{{Baibhav} \& {Berti}}{{Baibhav} \&
  {Berti}}{2019}]{Baibhav2018b}
{Baibhav} V.,  {Berti} E.,  2019, \mn@doi [\prd] {10.1103/PhysRevD.99.024005},
  \href {https://ui.adsabs.harvard.edu/abs/2019PhRvD..99b4005B} {99, 024005}

\bibitem[\protect\citeauthoryear{Baibhav, Berti, Cardoso  \& Khanna}{Baibhav
  et~al.}{2018}]{Baibhav2018a}
Baibhav V.,  Berti E.,  Cardoso V.,   Khanna G.,  2018, \mn@doi [\prd]
  {10.1103/PhysRevD.97.044048}, 97, 44048

\bibitem[\protect\citeauthoryear{Barausse, Bellovary, Berti, Holley-Bockelmann
  K.~Farris, Sathyaprakash  \& Sesana}{Barausse et~al.}{2015}]{Barausse2015}
Barausse E.,  Bellovary J.,  Berti E.,  Holley-Bockelmann K.~Farris B.,
  Sathyaprakash B.,   Sesana A.,  2015, J. Phys.: Conf. Ser., 610, 12001

\bibitem[\protect\citeauthoryear{Begelman, Blandford  \& Rees}{Begelman
  et~al.}{1980}]{Begelman1980}
Begelman M.~C.,  Blandford R.~D.,   Rees M.~J.,  1980, \mn@doi [\nat]
  {10.1038/287307a0}, 287, 307

\bibitem[\protect\citeauthoryear{Begelman, Volonteri  \& Rees}{Begelman
  et~al.}{2006}]{Begelman2006}
Begelman M.~C.,  Volonteri M.,   Rees M.~J.,  2006, \mn@doi [\mnras]
  {10.1111/j.1365-2966.2006.10467.x}, 370, 289

\bibitem[\protect\citeauthoryear{Bellovary, Cleary, Munshi, Tremmel,
  Christensen, Brooks  \& Quinn}{Bellovary et~al.}{2019}]{Bellovary2019}
Bellovary J.~M.,  Cleary C.~E.,  Munshi F.,  Tremmel M.,  Christensen C.~R.,
  Brooks A.,   Quinn T.~R.,  2019, \mn@doi [\mnras] {10.1093/mnras/sty2842},
  482, 2913

\bibitem[\protect\citeauthoryear{Bender \& Hils}{Bender \&
  Hils}{1997}]{HillsBender1997}
Bender P.~L.,  Hils D.,  1997, \mn@doi [Classical and Quantum Gravity]
  {10.1088/0264-9381/14/6/008}, 14, 1439

\bibitem[\protect\citeauthoryear{Berti, Cardoso  \& Will}{Berti
  et~al.}{2006}]{Berti2006}
Berti E.,  Cardoso V.,   Will C.~M.,  2006, \mn@doi [\prd]
  {10.1103/PhysRevD.73.064030}, 73, 64030

\bibitem[\protect\citeauthoryear{Berti, Sesana, Barausse, Cardoso  \&
  Belczynski}{Berti et~al.}{2016}]{Berti2016}
Berti E.,  Sesana A.,  Barausse E.,  Cardoso V.,   Belczynski K.,  2016,
  \mn@doi [Phys. Rev. Lett.] {10.1103/PhysRevLett.117.101102}, 117, 101102

\bibitem[\protect\citeauthoryear{Binney \& Tremaine}{Binney \&
  Tremaine}{1987}]{BinneyTremaineGalacticDynamics}
Binney J.,  Tremaine S.,  1987, {Galactic dynamics}

\bibitem[\protect\citeauthoryear{Blanchet}{Blanchet}{2014}]{Blanchet2014}
Blanchet L.,  2014, \mn@doi [Living Reviews in Relativity]
  {10.12942/lrr-2014-2}, 17, 2

\bibitem[\protect\citeauthoryear{Blecha et~al.,}{Blecha
  et~al.}{2016}]{Blecha2016}
Blecha L.,  et~al., 2016, \mn@doi [\mnras] {10.1093/mnras/stv2646}, 456, 961

\bibitem[\protect\citeauthoryear{Boehle et~al.,}{Boehle
  et~al.}{2016}]{BoehleGhez2016}
Boehle A.,  et~al., 2016, \mn@doi [\apj] {10.3847/0004-637X/830/1/17}, 830, 17

\bibitem[\protect\citeauthoryear{{Bogdanovi{\'c}}}{{Bogdanovi{\'c}}}{2015}]{Bogdanovic2015}
{Bogdanovi{\'c}} T.,  2015, in Gravitational Wave Astrophysics. p.~103
  (\mn@eprint {arXiv} {1406.5193}), \mn@doi{10.1007/978-3-319-10488-1_9}

\bibitem[\protect\citeauthoryear{{Bonetti}, {Sesana}, {Haardt}, {Barausse}  \&
  {Colpi}}{{Bonetti} et~al.}{2019}]{Bonetti2019}
{Bonetti} M.,  {Sesana} A.,  {Haardt} F.,  {Barausse} E.,   {Colpi} M.,  2019,
  \mn@doi [\mnras] {10.1093/mnras/stz903}, \href
  {https://ui.adsabs.harvard.edu/abs/2019MNRAS.486.4044B} {486, 4044}

\bibitem[\protect\citeauthoryear{{Burke-Spolaor}}{{Burke-Spolaor}}{2013}]{Burke-Spolaor2013}
{Burke-Spolaor} S.,  2013, \mn@doi [Classical and Quantum Gravity]
  {10.1088/0264-9381/30/22/224013}, \href
  {https://ui.adsabs.harvard.edu/abs/2013CQGra..30v4013B} {30, 224013}

\bibitem[\protect\citeauthoryear{Chandrasekhar}{Chandrasekhar}{1943}]{Chandrasekhar1943}
Chandrasekhar S.,  1943, \mn@doi [\apj] {10.1086/144517}, 97, 255

\bibitem[\protect\citeauthoryear{Charisi, Bartos, Haiman, Price-Whelan, Graham,
  Bellm, Laher  \& Márka}{Charisi et~al.}{2016}]{Charisi2016}
Charisi M.,  Bartos I.,  Haiman Z.,  Price-Whelan A.~M.,  Graham M.~J.,  Bellm
  E.~C.,  Laher R.~R.,   Márka S.,  2016, \mn@doi [\mnras]
  {10.1093/mnras/stw1838}, 463, 2145–2171

\bibitem[\protect\citeauthoryear{Davies, Miller  \& Bellovary}{Davies
  et~al.}{2011}]{Davies2011}
Davies M.~B.,  Miller M.~C.,   Bellovary J.~M.,  2011, \mn@doi [\apjl]
  {10.1088/2041-8205/740/2/L42}, 740, L42

\bibitem[\protect\citeauthoryear{Devecchi \& Volonteri}{Devecchi \&
  Volonteri}{2009}]{Devecchi2009}
Devecchi B.,  Volonteri M.,  2009, \mn@doi [\apj]
  {10.1088/0004-637X/694/1/302}, 694, 302

\bibitem[\protect\citeauthoryear{Dosopoulou \& Antonini}{Dosopoulou \&
  Antonini}{2017}]{Dosopoulou2017}
Dosopoulou F.,  Antonini F.,  2017, \mn@doi [\apj] {10.3847/1538-4357/aa6b58},
  840, 31

\bibitem[\protect\citeauthoryear{Dunn, Bellovary, Holley-Bockelmann,
  Christensen  \& Quinn}{Dunn et~al.}{2018}]{Dunn2018}
Dunn G.,  Bellovary J.,  Holley-Bockelmann K.,  Christensen C.,   Quinn T.,
  2018, \mn@doi [\apj] {10.3847/1538-4357/aac7c2}, 861, 39

\bibitem[\protect\citeauthoryear{Fan et~al.,}{Fan et~al.}{2001a}]{Fan2001b}
Fan X.,  et~al., 2001a, \mn@doi [\aj] {10.1086/318033}, 121, 54

\bibitem[\protect\citeauthoryear{Fan et~al.,}{Fan et~al.}{2001b}]{Fan2001a}
Fan X.,  et~al., 2001b, \mn@doi [\aj] {10.1086/324111}, 122, 2833

\bibitem[\protect\citeauthoryear{Fan et~al.,}{Fan et~al.}{2006}]{Fan2006}
Fan X.,  et~al., 2006, \mn@doi [\aj] {10.1086/500296}, 131, 1203

\bibitem[\protect\citeauthoryear{Finn \& Thorne}{Finn \&
  Thorne}{2000}]{Finn2000}
Finn L.~S.,  Thorne K.~S.,  2000, \mn@doi [\prd] {10.1103/PhysRevD.62.124021},
  62, 124021

\bibitem[\protect\citeauthoryear{{Fontecilla}, {Haiman}  \&
  {Cuadra}}{{Fontecilla} et~al.}{2019}]{Fontecilla2019}
{Fontecilla} C.,  {Haiman} Z.,   {Cuadra} J.,  2019, \mn@doi [\mnras]
  {10.1093/mnras/sty2972}, \href
  {http://adsabs.harvard.edu/abs/2019MNRAS.482.4383F} {482, 4383}

\bibitem[\protect\citeauthoryear{Frank \& Rees}{Frank \&
  Rees}{1976}]{Frank1976}
Frank J.,  Rees M.~J.,  1976, \mn@doi [\mnras] {10.1093/mnras/176.3.633}, 176,
  633

\bibitem[\protect\citeauthoryear{Fryer, Woosley  \& Heger}{Fryer
  et~al.}{2001}]{Fryer2001}
Fryer C.~L.,  Woosley S.~E.,   Heger A.,  2001, \mn@doi [\apj]
  {10.1086/319719}, 550, 372

\bibitem[\protect\citeauthoryear{Gair, Vallisneri, Larson  \& Baker}{Gair
  et~al.}{2013}]{Gair2013}
Gair J.~R.,  Vallisneri M.,  Larson S.~L.,   Baker J.~G.,  2013, \mn@doi
  [Living Reviews in Relativity] {10.12942/lrr-2013-7}, 16, 7

\bibitem[\protect\citeauthoryear{Genel et~al.,}{Genel et~al.}{2014}]{Genel2014}
Genel S.,  et~al., 2014, \mn@doi [\mnras] {10.1093/mnras/stu1654}, 445, 175

\bibitem[\protect\citeauthoryear{Graham et~al.,}{Graham
  et~al.}{2015}]{Graham2015a}
Graham M.~J.,  et~al., 2015, \mn@doi [\mnras] {10.1093/mnras/stv1726}, 453,
  1562

\bibitem[\protect\citeauthoryear{Habouzit, Volonteri, Latif, Dubois  \&
  Peirani}{Habouzit et~al.}{2016}]{Habouzit2016}
Habouzit M.,  Volonteri M.,  Latif M.,  Dubois Y.,   Peirani S.,  2016, \mn@doi
  [\mnras] {10.1093/mnras/stw1924}, 463, 529

\bibitem[\protect\citeauthoryear{Haiman, Abel  \& Rees}{Haiman
  et~al.}{2000}]{Haiman2000}
Haiman Z.,  Abel T.,   Rees M.~J.,  2000, \mn@doi [\apj] {10.1086/308723}, 534,
  11

\bibitem[\protect\citeauthoryear{Haiman, Kocsis  \& Menou}{Haiman
  et~al.}{2009}]{Haiman2009}
Haiman Z.,  Kocsis B.,   Menou K.,  2009, \mn@doi [\apj]
  {10.1088/0004-637X/700/2/1952}, 700, 1952

\bibitem[\protect\citeauthoryear{Heger, Fryer, Woosley, Langer  \&
  Hartmann}{Heger et~al.}{2003}]{Heger2003}
Heger A.,  Fryer C.~L.,  Woosley S.~E.,  Langer N.,   Hartmann D.~H.,  2003,
  \mn@doi [\apj] {10.1086/375341}, 591, 288

\bibitem[\protect\citeauthoryear{Hinshaw et~al.,}{Hinshaw
  et~al.}{2013}]{Hinshaw2012}
Hinshaw G.,  et~al., 2013, \mn@doi [\apjs] {10.1088/0067-0049/208/2/19}, 208,
  19

\bibitem[\protect\citeauthoryear{Hiscock, Larson, Routzahn  \& Kulick}{Hiscock
  et~al.}{2000}]{Hiscock2000}
Hiscock W.~A.,  Larson S.~L.,  Routzahn J.~R.,   Kulick B.,  2000, \mn@doi
  [\apjl] {10.1086/312867}, 540, L5

\bibitem[\protect\citeauthoryear{Holz \& Hughes}{Holz \&
  Hughes}{2005}]{Holz2005}
Holz D.~E.,  Hughes S.~A.,  2005, \mn@doi [\apj] {10.1086/431341}, 629, 15

\bibitem[\protect\citeauthoryear{Hunter}{Hunter}{2007}]{Matplotlib}
Hunter J.~D.,  2007, \mn@doi [Computing In Science {\&} Engineering]
  {10.1109/MCSE.2007.55}, 9, 90

\bibitem[\protect\citeauthoryear{Husa, Khan, Hannam, P{\"{u}}rrer, Ohme,
  Forteza  \& Boh{\'{e}}}{Husa et~al.}{2016}]{Husa2016}
Husa S.,  Khan S.,  Hannam M.,  P{\"{u}}rrer M.,  Ohme F.,  Forteza X.~J.,
  Boh{\'{e}} A.,  2016, \mn@doi [\prd] {10.1103/PhysRevD.93.044006}, 93, 44006

\bibitem[\protect\citeauthoryear{Jones, Oliphant, Peterson  et~al.}{Jones
  et~al.}{2001}]{scipy}
Jones E.,  Oliphant T.,  Peterson P.,   et~al., 2001, {SciPy}: Open source
  scientific tools for {Python}, \url {http://www.scipy.org/}

\bibitem[\protect\citeauthoryear{Katz \& Larson}{Katz \&
  Larson}{2019}]{Katz2018}
Katz M.~L.,  Larson S.~L.,  2019, \mn@doi [\mnras] {10.1093/mnras/sty3321},
  483, 3108

\bibitem[\protect\citeauthoryear{Katz, Sijacki  \& Haehnelt}{Katz
  et~al.}{2015}]{Katz2015}
Katz H.,  Sijacki D.,   Haehnelt M.~G.,  2015, \mn@doi [\mnras]
  {10.1093/mnras/stv1048}, 451, 2352

\bibitem[\protect\citeauthoryear{Kelley, Blecha  \& Hernquist}{Kelley
  et~al.}{2017a}]{PTA-illustris}
Kelley L.~Z.,  Blecha L.,   Hernquist L.,  2017a, \mn@doi [\mnras]
  {10.1093/mnras/stw2452}, 464, 3131

\bibitem[\protect\citeauthoryear{Kelley, Blecha, Hernquist, Sesana  \&
  Taylor}{Kelley et~al.}{2017b}]{Kelley2017b}
Kelley L.~Z.,  Blecha L.,  Hernquist L.,  Sesana A.,   Taylor S.~R.,  2017b,
  \mn@doi [\mnras] {10.1093/mnras/stx1638}, 471, 4508

\bibitem[\protect\citeauthoryear{{Kelley}, {Blecha}, {Hernquist}, {Sesana}  \&
  {Taylor}}{{Kelley} et~al.}{2018}]{Kelley2018}
{Kelley} L.~Z.,  {Blecha} L.,  {Hernquist} L.,  {Sesana} A.,   {Taylor} S.~R.,
  2018, \mn@doi [\mnras] {10.1093/mnras/sty689}, \href
  {https://ui.adsabs.harvard.edu/abs/2018MNRAS.477..964K} {477, 964}

\bibitem[\protect\citeauthoryear{Khan, Just  \& Merritt}{Khan
  et~al.}{2011}]{Khan2011}
Khan F.~M.,  Just A.,   Merritt D.,  2011, \mn@doi [\apj]
  {10.1088/0004-637X/732/2/89}, 732, 89

\bibitem[\protect\citeauthoryear{Khan, Husa, Hannam, Ohme, P{\"{u}}rrer,
  Forteza  \& Boh{\'{e}}}{Khan et~al.}{2016}]{Khan2016}
Khan S.,  Husa S.,  Hannam M.,  Ohme F.,  P{\"{u}}rrer M.,  Forteza X.~J.,
  Boh{\'{e}} A.,  2016, \mn@doi [\prd] {10.1103/PhysRevD.93.044007}, 93, 44007

\bibitem[\protect\citeauthoryear{Klein et~al.,}{Klein et~al.}{2016}]{Klein2016}
Klein A.,  et~al., 2016, \mn@doi [\prd] {10.1103/PhysRevD.93.024003}, 93, 24003

\bibitem[\protect\citeauthoryear{Kormendy \& Richstone}{Kormendy \&
  Richstone}{1995}]{Kormendy1995}
Kormendy J.,  Richstone D.,  1995, \mn@doi [\araa]
  {10.1146/annurev.aa.33.090195.003053}, 33, 581

\bibitem[\protect\citeauthoryear{Larson, Hiscock  \& Hellings}{Larson
  et~al.}{2000}]{Larson2000}
Larson S.~L.,  Hiscock W.~A.,   Hellings R.~W.,  2000, \mn@doi [\prd]
  {10.1103/PhysRevD.62.062001}, 62, 62001

\bibitem[\protect\citeauthoryear{Latif, Schleicher, Schmidt  \& Niemeyer}{Latif
  et~al.}{2013}]{Latif2013}
Latif M.~A.,  Schleicher D.~R.~G.,  Schmidt W.,   Niemeyer J.~C.,  2013,
  \mn@doi [\mnras] {10.1093/mnras/stt1786}, 436, 2989

\bibitem[\protect\citeauthoryear{Lauer, Tremaine, Richstone  \& Faber}{Lauer
  et~al.}{2007}]{Lauer2007}
Lauer T.~R.,  Tremaine S.,  Richstone D.,   Faber S.~M.,  2007, \mn@doi [\apj]
  {10.1086/522083}, 670, 249

\bibitem[\protect\citeauthoryear{Lemons, Reines, Plotkin, Gallo  \&
  Greene}{Lemons et~al.}{2015}]{Lemons2015}
Lemons S.~M.,  Reines A.~E.,  Plotkin R.~M.,  Gallo E.,   Greene J.~E.,  2015,
  \mn@doi [\apj] {10.1088/0004-637X/805/1/12}, 805, 12

\bibitem[\protect\citeauthoryear{Lightman \& Shapiro}{Lightman \&
  Shapiro}{1977}]{Lightman1977}
Lightman A.~P.,  Shapiro S.~L.,  1977, \mn@doi [\apj] {10.1086/154925}, 211,
  244

\bibitem[\protect\citeauthoryear{Lin et~al.,}{Lin et~al.}{2018}]{Lin2018}
Lin D.,  et~al., 2018, \mn@doi [Nature Astronomy] {10.1038/s41550-018-0493-1},
  2, 656

\bibitem[\protect\citeauthoryear{Liu et~al.,}{Liu et~al.}{2016}]{Liu2016}
Liu T.,  et~al., 2016, \mn@doi [\apj] {10.3847/0004-637x/833/1/6}, 833, 6

\bibitem[\protect\citeauthoryear{Loeb \& Rasio}{Loeb \& Rasio}{1994}]{Loeb1994}
Loeb A.,  Rasio F.~A.,  1994, \mn@doi [\apj] {10.1086/174548}, 432, 52

\bibitem[\protect\citeauthoryear{Magorrian et~al.,}{Magorrian
  et~al.}{1998}]{Magorrian1998}
Magorrian J.,  et~al., 1998, \mn@doi [\aj] {10.1086/300353}, 115, 2285

\bibitem[\protect\citeauthoryear{McAlpine et~al.,}{McAlpine
  et~al.}{2016}]{EAGLE-release}
McAlpine S.,  et~al., 2016, \mn@doi [Astronomy and Computing]
  {10.1016/j.ascom.2016.02.004}, 15, 72

\bibitem[\protect\citeauthoryear{McConnell \& Ma}{McConnell \&
  Ma}{2013}]{McConnell2013}
McConnell N.~J.,  Ma C.-P.,  2013, \mn@doi [\apj]
  {10.1088/0004-637X/764/2/184}, 764, 184

\bibitem[\protect\citeauthoryear{Merritt}{Merritt}{2013}]{Merritt2013}
Merritt D.,  2013, {Dynamics and Evolution of Galactic Nuclei}

\bibitem[\protect\citeauthoryear{Merritt \& Ferrarese}{Merritt \&
  Ferrarese}{2001}]{Merritt2001}
Merritt D.,  Ferrarese L.,  2001, \mn@doi [\mnras]
  {10.1046/j.1365-8711.2001.04165.x}, 320, L30

\bibitem[\protect\citeauthoryear{Merritt \& Milosavljevi{\'{c}}}{Merritt \&
  Milosavljevi{\'{c}}}{2005}]{Merritt2005}
Merritt D.,  Milosavljevi{\'{c}} M.,  2005, \mn@doi [Living Reviews in
  Relativity] {10.12942/lrr-2005-8}, 8, 8

\bibitem[\protect\citeauthoryear{Merritt, Schnittman  \& Komossa}{Merritt
  et~al.}{2009}]{Merritt2009}
Merritt D.,  Schnittman J.~D.,   Komossa S.,  2009, \mn@doi [\apj]
  {10.1088/0004-637X/699/2/1690}, 699, 1690

\bibitem[\protect\citeauthoryear{Miller}{Miller}{2007}]{Miller2007}
Miller J.~M.,  2007, \mn@doi [\araa] {10.1146/annurev.astro.45.051806.110555},
  45, 441

\bibitem[\protect\citeauthoryear{Mirza, Tahir, Khan, Holley-Bockelmann, Baig,
  Berczik  \& Chishtie}{Mirza et~al.}{2017}]{Mirza2017}
Mirza M.~A.,  Tahir A.,  Khan F.~M.,  Holley-Bockelmann H.,  Baig A.~M.,
  Berczik P.,   Chishtie F.,  2017, \mn@doi [\mnras] {10.1093/mnras/stx1248},
  470, 940

\bibitem[\protect\citeauthoryear{Moody, Shi  \& Stone}{Moody
  et~al.}{2019}]{Moody2019}
Moody M. S.~L.,  Shi J.-M.,   Stone J.~M.,  2019, \mn@doi [\apj]
  {10.3847/1538-4357/ab09ee}, 875, 66

\bibitem[\protect\citeauthoryear{Moore, Cole  \& Berry}{Moore
  et~al.}{2015}]{Moore2015}
Moore C.~J.,  Cole R.~H.,   Berry C. P.~L.,  2015, Classical and Quantum
  Gravity, 32, 15014

\bibitem[\protect\citeauthoryear{Moran, Shahinyan, Sugarman, V{\'{e}}lez  \&
  Eracleous}{Moran et~al.}{2014}]{Moran2014}
Moran E.~C.,  Shahinyan K.,  Sugarman H.~R.,  V{\'{e}}lez D.~O.,   Eracleous
  M.,  2014, \mn@doi [\aj] {10.1088/0004-6256/148/6/136}, 148, 136

\bibitem[\protect\citeauthoryear{Mortlock et~al.,}{Mortlock
  et~al.}{2011}]{Mortlock2011}
Mortlock D.~J.,  et~al., 2011, \mn@doi [\nat] {10.1038/nature10159}, 474, 616

\bibitem[\protect\citeauthoryear{{Mu{\~n}oz}, {Miranda}  \& {Lai}}{{Mu{\~n}oz}
  et~al.}{2019}]{Munoz2019}
{Mu{\~n}oz} D.~J.,  {Miranda} R.,   {Lai} D.,  2019, \mn@doi [\apj]
  {10.3847/1538-4357/aaf867}, \href
  {https://ui.adsabs.harvard.edu/abs/2019ApJ...871...84M} {871, 84}

\bibitem[\protect\citeauthoryear{Nelson et~al.,}{Nelson
  et~al.}{2015}]{Nelson2015}
Nelson D.,  et~al., 2015, \mn@doi [Astronomy and Computing]
  {10.1016/j.ascom.2015.09.003}, 13, 12

\bibitem[\protect\citeauthoryear{{Nguyen} et~al.,}{{Nguyen}
  et~al.}{2018}]{Nguyen2018}
{Nguyen} D.~D.,  et~al., 2018, \mn@doi [\apj] {10.3847/1538-4357/aabe28}, \href
  {https://ui.adsabs.harvard.edu/abs/2018ApJ...858..118N} {858, 118}

\bibitem[\protect\citeauthoryear{{Nguyen} et~al.,}{{Nguyen}
  et~al.}{2019}]{Nguyen2019}
{Nguyen} D.~D.,  et~al., 2019, \mn@doi [\apj] {10.3847/1538-4357/aafe7a}, \href
  {https://ui.adsabs.harvard.edu/abs/2019ApJ...872..104N} {872, 104}

\bibitem[\protect\citeauthoryear{Omukai, Schneider  \& Haiman}{Omukai
  et~al.}{2008}]{Omukai2008}
Omukai K.,  Schneider R.,   Haiman Z.,  2008, \mn@doi [\apj] {10.1086/591636},
  686, 801

\bibitem[\protect\citeauthoryear{Pardo et~al.,}{Pardo et~al.}{2016}]{Pardo2016}
Pardo K.,  et~al., 2016, \mn@doi [\apj] {10.3847/0004-637X/831/2/203}, 831, 203

\bibitem[\protect\citeauthoryear{Peters \& Mathews}{Peters \&
  Mathews}{1963}]{Peters1963}
Peters P.~C.,  Mathews J.,  1963, \mn@doi [Physical Review]
  {10.1103/PhysRev.131.435}, 131, 435

\bibitem[\protect\citeauthoryear{Petiteau, Babak  \& Sesana}{Petiteau
  et~al.}{2011}]{Petiteau2011}
Petiteau A.,  Babak S.,   Sesana A.,  2011, \mn@doi [\apj]
  {10.1088/0004-637X/732/2/82}, 732, 82

\bibitem[\protect\citeauthoryear{Plowman, Jacobs, Hellings, Larson  \&
  Tsuruta}{Plowman et~al.}{2010}]{Plowman2010}
Plowman J.~E.,  Jacobs D.~C.,  Hellings R.~W.,  Larson S.~L.,   Tsuruta S.,
  2010, \mn@doi [\mnras] {10.1111/j.1365-2966.2009.15853.x}, 401, 2706

\bibitem[\protect\citeauthoryear{Plowman, Hellings  \& Tsuruta}{Plowman
  et~al.}{2011}]{Plowman2011}
Plowman J.~E.,  Hellings R.~W.,   Tsuruta S.,  2011, \mn@doi [\mnras]
  {10.1111/j.1365-2966.2011.18703.x}, 415, 333

\bibitem[\protect\citeauthoryear{{Porter} \& {Sesana}}{{Porter} \&
  {Sesana}}{2010}]{Porter2010}
{Porter} E.~K.,  {Sesana} A.,  2010, arXiv e-prints, \href
  {https://ui.adsabs.harvard.edu/abs/2010arXiv1005.5296P} {p. arXiv:1005.5296}

\bibitem[\protect\citeauthoryear{Quinlan}{Quinlan}{1996}]{Quinlan1996}
Quinlan G.~D.,  1996, \mn@doi [\na] {10.1016/S1384-1076(96)00018-8}, 1, 255

\bibitem[\protect\citeauthoryear{Quinlan \& Hernquist}{Quinlan \&
  Hernquist}{1997}]{Quinlan1997}
Quinlan G.~D.,  Hernquist L.,  1997, \mn@doi [\na]
  {10.1016/S1384-1076(97)00039-0}, 2, 533

\bibitem[\protect\citeauthoryear{Rasskazov \& Merritt}{Rasskazov \&
  Merritt}{2017}]{Rasskazov2}
Rasskazov A.,  Merritt D.,  2017, \mn@doi [\apj] {10.3847/1538-4357/aa6188},
  837, 135

\bibitem[\protect\citeauthoryear{Reines, Greene  \& Geha}{Reines
  et~al.}{2013}]{Reines2013}
Reines A.~E.,  Greene J.~E.,   Geha M.,  2013, \mn@doi [\apj]
  {10.1088/0004-637X/775/2/116}, 775, 116

\bibitem[\protect\citeauthoryear{Reynolds}{Reynolds}{2013}]{Reynolds2013}
Reynolds C.~S.,  2013, \mn@doi [Classical and Quantum Gravity]
  {10.1088/0264-9381/30/24/244004}, 30, 244004

\bibitem[\protect\citeauthoryear{Robson \& Cornish}{Robson \&
  Cornish}{2017}]{Robson2017}
Robson T.,  Cornish N.,  2017, preprint

\bibitem[\protect\citeauthoryear{{Robson}, {Cornish}  \& {Liu}}{{Robson}
  et~al.}{2019}]{Robson2019}
{Robson} T.,  {Cornish} N.~J.,   {Liu} C.,  2019, \mn@doi [Classical and
  Quantum Gravity] {10.1088/1361-6382/ab1101}, \href
  {https://ui.adsabs.harvard.edu/abs/2019CQGra..36j5011R} {36, 105011}

\bibitem[\protect\citeauthoryear{Rodriguez-Gomez et~al.,}{Rodriguez-Gomez
  et~al.}{2015}]{Rodriguez-Gomez2015}
Rodriguez-Gomez V.,  et~al., 2015, \mn@doi [\mnras] {10.1093/mnras/stv264},
  449, 49

\bibitem[\protect\citeauthoryear{Roedig, Dotti, Sesana, Cuadra  \&
  Colpi}{Roedig et~al.}{2011}]{Roedig2011}
Roedig C.,  Dotti M.,  Sesana A.,  Cuadra J.,   Colpi M.,  2011, \mn@doi
  [\mnras] {10.1111/j.1365-2966.2011.18927.x}, 415, 3033

\bibitem[\protect\citeauthoryear{Salcido, Bower, Theuns, McAlpine, Schaller,
  Crain, Schaye  \& Regan}{Salcido et~al.}{2016}]{Salcido2016}
Salcido J.,  Bower R.~G.,  Theuns T.,  McAlpine S.,  Schaller M.,  Crain R.~A.,
   Schaye J.,   Regan J.,  2016, \mn@doi [\mnras] {10.1093/mnras/stw2048}, 463,
  870

\bibitem[\protect\citeauthoryear{Sartori, Schawinski, Treister, Trakhtenbrot,
  Koss, Shirazi  \& Oh}{Sartori et~al.}{2015}]{Sartori2015}
Sartori L.~F.,  Schawinski K.,  Treister E.,  Trakhtenbrot B.,  Koss M.,
  Shirazi M.,   Oh K.,  2015, \mn@doi [\mnras] {10.1093/mnras/stv2238}, 454,
  3722

\bibitem[\protect\citeauthoryear{Satyapal, Secrest, McAlpine, Ellison, Fischer
  \& Rosenberg}{Satyapal et~al.}{2014}]{Satyapal2014}
Satyapal S.,  Secrest N.~J.,  McAlpine W.,  Ellison S.~L.,  Fischer J.,
  Rosenberg J.~L.,  2014, \mn@doi [\apj] {10.1088/0004-637x/784/2/113}, 784,
  113

\bibitem[\protect\citeauthoryear{{Schulze} \& {Wisotzki}}{{Schulze} \&
  {Wisotzki}}{2011}]{Schulze2011}
{Schulze} A.,  {Wisotzki} L.,  2011, \mn@doi [\aap]
  {10.1051/0004-6361/201117564}, \href
  {https://ui.adsabs.harvard.edu/abs/2011A&A...535A..87S} {535, A87}

\bibitem[\protect\citeauthoryear{Schutz}{Schutz}{1986}]{Schutz1986}
Schutz B.~F.,  1986, \mn@doi [\nat] {10.1038/323310a0}, 323, 310

\bibitem[\protect\citeauthoryear{Sesana}{Sesana}{2010}]{Sesana2010}
Sesana A.,  2010, \mn@doi [\apj] {10.1088/0004-637X/719/1/851}, 719, 851

\bibitem[\protect\citeauthoryear{Sesana, Haardt, Madau  \& Volonteri}{Sesana
  et~al.}{2004}]{Sesana2004}
Sesana A.,  Haardt F.,  Madau P.,   Volonteri M.,  2004, \mn@doi [\apj]
  {10.1086/422185}, 611, 623

\bibitem[\protect\citeauthoryear{{Sesana}, {Haardt}, {Madau}  \&
  {Volonteri}}{{Sesana} et~al.}{2005}]{Sesana2005}
{Sesana} A.,  {Haardt} F.,  {Madau} P.,   {Volonteri} M.,  2005, \mn@doi [\apj]
  {10.1086/428492}, \href
  {https://ui.adsabs.harvard.edu/abs/2005ApJ...623...23S} {623, 23}

\bibitem[\protect\citeauthoryear{Sesana, Haardt  \& Madau}{Sesana
  et~al.}{2006}]{sesana2006}
Sesana A.,  Haardt F.,   Madau P.,  2006, \mn@doi [\apj] {10.1086/507596}, 651,
  392

\bibitem[\protect\citeauthoryear{Sesana, Gair, Berti  \& Volonteri}{Sesana
  et~al.}{2011}]{Sesana2011}
Sesana A.,  Gair J.,  Berti E.,   Volonteri M.,  2011, \mn@doi [\prd]
  {10.1103/PhysRevD.83.044036}, 83, 44036

\bibitem[\protect\citeauthoryear{Shakura \& Sunyaev}{Shakura \&
  Sunyaev}{1973}]{Shakura1973}
Shakura N.~I.,  Sunyaev R.~A.,  1973, \aap, 500, 33

\bibitem[\protect\citeauthoryear{Shankar et~al.,}{Shankar
  et~al.}{2016}]{Shankar2016}
Shankar F.,  et~al., 2016, \mn@doi [\mnras] {10.1093/mnras/stw678}, 460, 3119

\bibitem[\protect\citeauthoryear{{Shapiro} \& {Teukolsky}}{{Shapiro} \&
  {Teukolsky}}{1986}]{Shapiro1986}
{Shapiro} S.~L.,  {Teukolsky} S.~A.,  1986, {Black Holes, White Dwarfs and
  Neutron Stars: The Physics of Compact Objects}

\bibitem[\protect\citeauthoryear{Shen, Greene, Strauss, Richards  \&
  Schneider}{Shen et~al.}{2008}]{Shen2008}
Shen Y.,  Greene J.~E.,  Strauss M.~A.,  Richards G.~T.,   Schneider D.~P.,
  2008, \mn@doi [\apj] {10.1086/587475}, 680, 169

\bibitem[\protect\citeauthoryear{Sijacki, Vogelsberger, Genel, Springel,
  Torrey, Snyder, Nelson  \& Hernquist}{Sijacki et~al.}{2015}]{Illustris-BHs}
Sijacki D.,  Vogelsberger M.,  Genel S.,  Springel V.,  Torrey P.,  Snyder
  G.~F.,  Nelson D.,   Hernquist L.,  2015, \mn@doi [\mnras]
  {10.1093/mnras/stv1340}, 452, 575

\bibitem[\protect\citeauthoryear{Soltan}{Soltan}{1982}]{Soltan1982}
Soltan A.,  1982, \mn@doi [\mnras] {10.1093/mnras/200.1.115}, 200, 115

\bibitem[\protect\citeauthoryear{Springel}{Springel}{2010}]{Springel2010}
Springel V.,  2010, \mn@doi [\mnras] {10.1111/j.1365-2966.2009.15715.x}, 401,
  791

\bibitem[\protect\citeauthoryear{Tanaka \& Haiman}{Tanaka \&
  Haiman}{2009}]{Tanaka2009}
Tanaka T.,  Haiman Z.,  2009, \mn@doi [\apj] {10.1088/0004-637X/696/2/1798},
  696, 1798

\bibitem[\protect\citeauthoryear{{Tang}, {MacFadyen}  \& {Haiman}}{{Tang}
  et~al.}{2017}]{Tang2017}
{Tang} Y.,  {MacFadyen} A.,   {Haiman} Z.,  2017, \mn@doi [\mnras]
  {10.1093/mnras/stx1130}, \href
  {http://adsabs.harvard.edu/abs/2017MNRAS.469.4258T} {469, 4258}

\bibitem[\protect\citeauthoryear{Torrey, Vogelsberger, Genel, Sijacki, Springel
   \& Hernquist}{Torrey et~al.}{2014}]{Torrey2014}
Torrey P.,  Vogelsberger M.,  Genel S.,  Sijacki D.,  Springel V.,   Hernquist
  L.,  2014, \mn@doi [\mnras] {10.1093/mnras/stt2295}, 438, 1985

\bibitem[\protect\citeauthoryear{Tremmel, Governato, Volonteri, Quinn  \&
  Pontzen}{Tremmel et~al.}{2018}]{Tremmel2018}
Tremmel M.,  Governato F.,  Volonteri M.,  Quinn T.~R.,   Pontzen A.,  2018,
  \mn@doi [\mnras] {10.1093/mnras/sty139}, 475, 4967

\bibitem[\protect\citeauthoryear{Vasiliev \& Merritt}{Vasiliev \&
  Merritt}{2013}]{Vasiliev2013}
Vasiliev E.,  Merritt D.,  2013, \mn@doi [\apj] {10.1088/0004-637X/774/1/87},
  774, 87

\bibitem[\protect\citeauthoryear{Vasiliev, Antonini  \& Merritt}{Vasiliev
  et~al.}{2014}]{Vasiliev2014}
Vasiliev E.,  Antonini F.,   Merritt D.,  2014, \mn@doi [\apj]
  {10.1088/0004-637X/785/2/163}, 785, 163

\bibitem[\protect\citeauthoryear{Vasiliev, Antonini  \& Merritt}{Vasiliev
  et~al.}{2015}]{Vasiliev2015}
Vasiliev E.,  Antonini F.,   Merritt D.,  2015, \mn@doi [\apj]
  {10.1088/0004-637X/810/1/49}, 810, 49

\bibitem[\protect\citeauthoryear{Vogelsberger, Genel, Sijacki, Torrey, Springel
   \& Hernquist}{Vogelsberger et~al.}{2013}]{Vogelsberger2013}
Vogelsberger M.,  Genel S.,  Sijacki D.,  Torrey P.,  Springel V.,   Hernquist
  L.,  2013, \mn@doi [\mnras] {10.1093/mnras/stt1789}, 436, 3031

\bibitem[\protect\citeauthoryear{Vogelsberger et~al.,}{Vogelsberger
  et~al.}{2014a}]{Vogelsberger2014a}
Vogelsberger M.,  et~al., 2014a, MNRAS, 444, 1518

\bibitem[\protect\citeauthoryear{Vogelsberger et~al.,}{Vogelsberger
  et~al.}{2014b}]{Vogelsberger2014b}
Vogelsberger M.,  et~al., 2014b, \mn@doi [\nat] {10.1038/nature13316}, 509, 177

\bibitem[\protect\citeauthoryear{Volonteri, Madau  \& Haardt}{Volonteri
  et~al.}{2003}]{Volonteri2003}
Volonteri M.,  Madau P.,   Haardt F.,  2003, \mn@doi [\apj] {10.1086/376722},
  593, 661

\bibitem[\protect\citeauthoryear{{Volonteri}, {Lodato}  \&
  {Natarajan}}{{Volonteri} et~al.}{2008}]{Volonteri2008}
{Volonteri} M.,  {Lodato} G.,   {Natarajan} P.,  2008, \mn@doi [\mnras]
  {10.1111/j.1365-2966.2007.12589.x}, \href
  {https://ui.adsabs.harvard.edu/abs/2008MNRAS.383.1079V} {383, 1079}

\bibitem[\protect\citeauthoryear{Walt, Colbert  \& Varoquaux}{Walt
  et~al.}{2011}]{Numpy}
Walt S. v.~d.,  Colbert S.~C.,   Varoquaux G.,  2011, \mn@doi [Computing in
  Science and Engg.] {10.1109/MCSE.2011.37}, 13, 22

\bibitem[\protect\citeauthoryear{{Wang} et~al.,}{{Wang}
  et~al.}{2019}]{Wang2019}
{Wang} H.-T.,  et~al., 2019, \mn@doi [\prd] {10.1103/PhysRevD.100.043003},
  \href {https://ui.adsabs.harvard.edu/abs/2019PhRvD.100d3003W} {100, 043003}

\bibitem[\protect\citeauthoryear{White \& Frenk}{White \&
  Frenk}{1991}]{White1991}
White S.~D.~M.,  Frenk C.~S.,  1991, \mn@doi [\apj] {10.1086/170483}, 379, 52

\bibitem[\protect\citeauthoryear{Yu}{Yu}{2002}]{Yu2002}
Yu Q.,  2002, \mn@doi [\mnras] {10.1046/j.1365-8711.2002.05242.x}, 331, 935

\bibitem[\protect\citeauthoryear{{eLISA Consortium} et~al.,}{{eLISA Consortium}
  et~al.}{2013}]{GravitationalUniverse}
{eLISA Consortium} et~al., 2013, arXiv e-prints, p. arXiv:1305.5720

\bibitem[\protect\citeauthoryear{{van Wassenhove}, {Volonteri}, {Walker}  \&
  {Gair}}{{van Wassenhove} et~al.}{2010}]{vanWassenhove2010}
{van Wassenhove} S.,  {Volonteri} M.,  {Walker} M.~G.,   {Gair} J.~R.,  2010,
  \mn@doi [\mnras] {10.1111/j.1365-2966.2010.17189.x}, \href
  {https://ui.adsabs.harvard.edu/abs/2010MNRAS.408.1139V} {408, 1139}

\makeatother
\end{thebibliography}

\appendix
\section{Effect of LISA Configuration on Rate Predictions}\label{sec:lisaconfig}

\begin{figure}
\begin{center}
\includegraphics[scale=0.33]{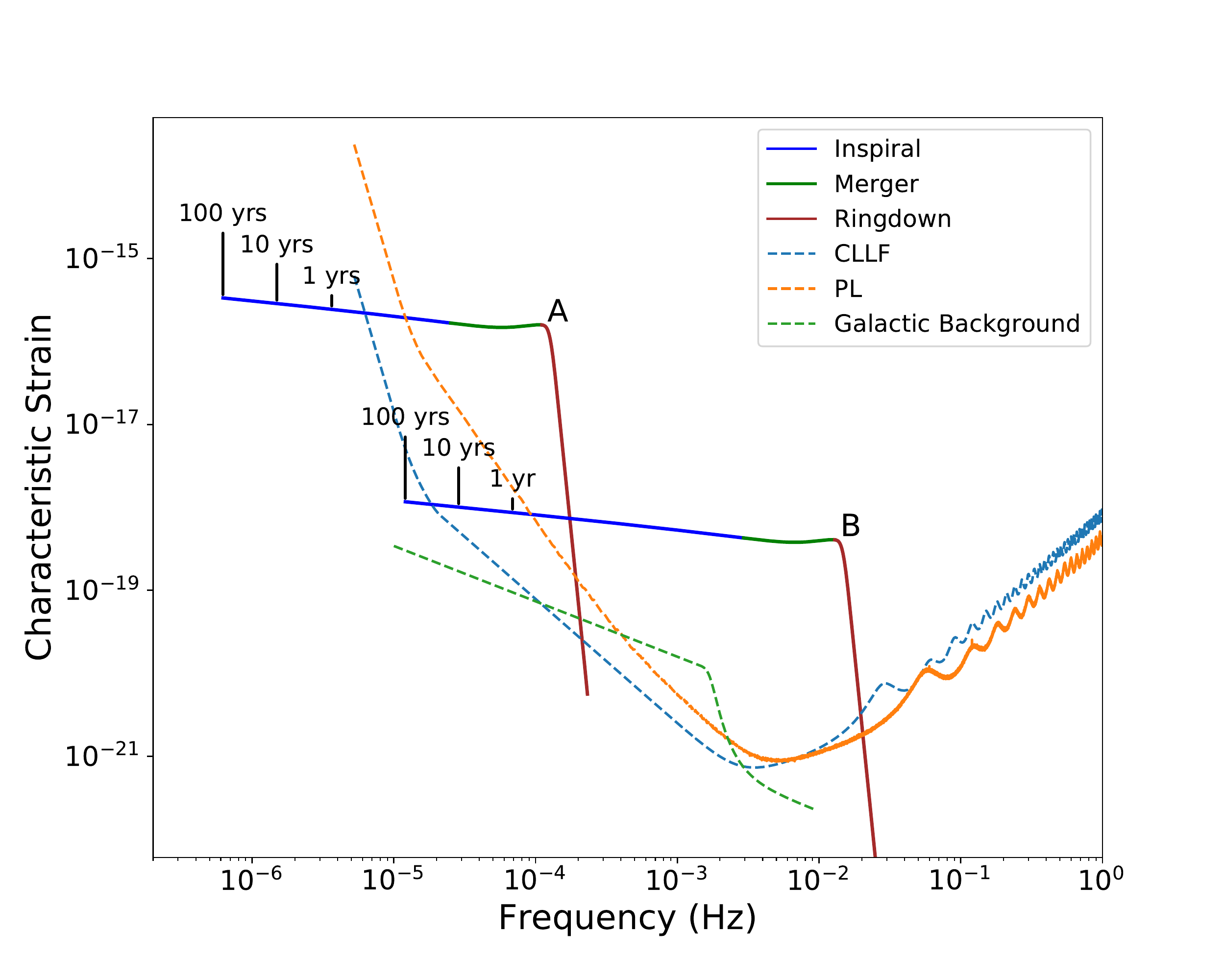}
\caption{This figure is the same as Figure \ref{fig:sensitivity_curves_and_signals}, but with the addition of the modified classic LISA curve (see below). Two examples of the characteristic strain, $h_c$, curves are shown here with solid lines. The blue, green, and red portions of the binary signals represent the construction we use for the inspiral, merger, and ringdown, respectively. Both examples show $a=0.8$ and $q=0.2$ for a signal beginning 100 years before merger. To plot these curves, we use $t_\text{st}=100$ yrs and $t_\text{end}=0$ so that we encapsulate 100 years of inspiral as well as the merger and ringdown. The times before merger are labeled above the strain curve for 100, 10, and 1 yrs before merger. Example A shows a binary of $M_T=10^8M_\odot$ and $z=0.75$. Example B shows $M_T=5\times10^5M_\odot$ and $z=2$. In addition to binary signals, the two sensitivity curves tested in this section are shown in characteristic strain of the noise, $h_N$. PL (dashed orange) is the curve proposed in \citet{LISAMissionProposal}. CLLF (dashed blue) is a modified version of the classic LISA curve \citep{Larson2000}. At low frequencies, the classic LISA curve has an unphysical constant slope. To correct for this, we move the PL low-frequency behavior to lower strains and spline it together with the original classic LISA curve. Additionally, the Galactic background noise we use is shown with a dashed green line.} \label{fig:sensitivity_curves_and_signals2}
\end{center}
\end{figure}

\begin{figure}
\begin{center}
\includegraphics[scale=0.4]{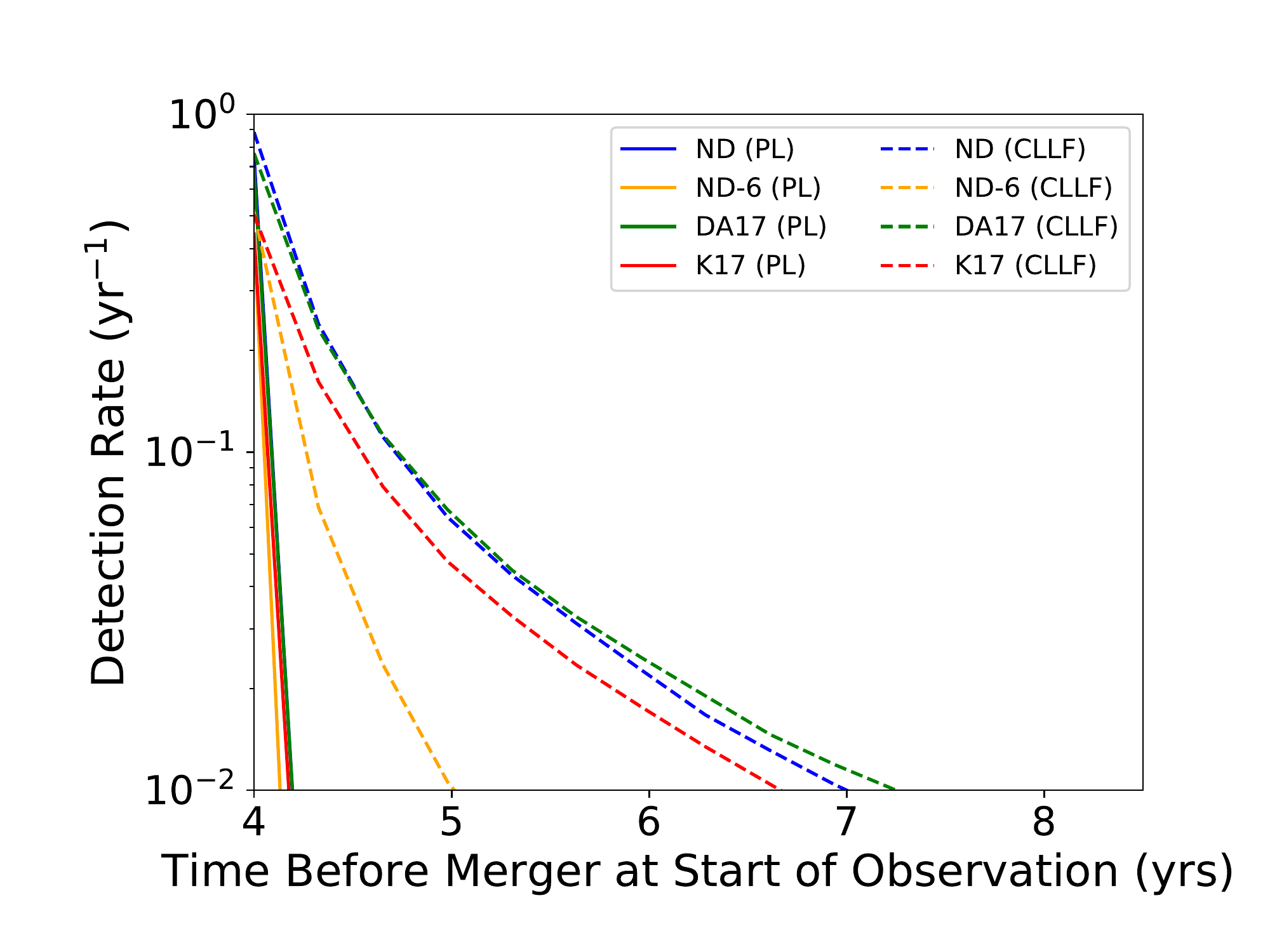}
\end{center}
\caption{The integral detection rate calculation for different values of $t_\text{st}$ is shown above. $t_\text{st}$ is set to the value on the horizontal axis. $T_\text{obs}$ is set to 4 yrs. Each binary inspiral model is then tested with both LISA configurations. The ``no delay'' (ND) model is shown in blue. ND-6, a subset of the ND model with $m_1,m_2\geq10^6M_\odot$, is shown in orange. DA17 \citep{Dosopoulou2017} and K17 \citep{PTA-illustris, Kelley2017b} models are shown in green and red, respectively. The PL (CLLF) LISA configuration is shown with solid (dashed) lines.}
\label{fig:time_to_merger}
\end{figure}

\begin{table*}
\centering
 \begin{tabular}{||c c c c c c c||} 
 \hline
 Prescription & PL & CLLF & PL & CLLF &  PL & CLLF \\ [0.5ex] 
 - & All Signal & All Signal & Ins Only & Ins Only &  Fraction of Ins Only & Fraction of Ins Only \\
 \hline
ND  & 0.79 & 1.08 & <0.01 & 0.19 & <1\% & 18\% \\
\textit{ND-6} & \textit{0.47} & \textit{0.57} & \textit{<0.01} & \textit{0.05} & \textit{<1\%} & \textit{9\%} \\
DA17 & 0.72 & 0.97 & <0.01 & 0.20 & <1\% & 21\% \\
K17 & 0.47 & 0.64 & <0.01 & 0.13  & <1\% & 20\% \\
 \hline
\end{tabular}
\caption{Monte Carlo results for the observed detection rate per year from our 10000 catalogs are shown above (see \ref{sec:montecarlo}). We focused our Monte Carlo calculation on the overall signal detection rate as well as sources only detected in their inspiral stage. The coalescence timescale prescriptions are listed in the first column: ND is the ``no delay'' model; ND-6 is a subset of the ND model with all constituent masses below $10^6M_\odot$ eliminated from consideration; DA17 is the binary inspiral model from \citet{Dosopoulou2017}; and K17 is the inspiral model from \citet{PTA-illustris, Kelley2017b}. The full signal detection rate with PL and CLLF are shown in the second and third columns, respectivley. Similary, inspiral-only detection rates are shown for PL and CLLF in the fourth and fifth columns, respectively. The last two columns show the fraction of the total detection rate contributed by inspiral-only signals using PL and CLLF. For our Monte Carlo results, the error in the predictions is approximately $\sqrt{1/N}=\sqrt{1/10000}=1\%$.}
\label{tb:monte_carlo_rates}
\end{table*}

\begin{table*}
\centering
 \begin{tabular}{||c c c c c c c c c c||} 
 \hline
  - & - & PL & CLLF & PL & CLLF &  PL & CLLF & PL & CLLF \\ [0.5ex] 
 Prescription & Merger Rate & All & All & Ins & Ins & MR & MR & BHS & BHS \\
 \hline
ND & 0.98 & 0.75 & 0.89 & 0.45 & 0.75 & 0.77 & 0.89 & 0.74 & 0.88 \\
\textit{ND-6} & \textit{0.57} & \textit{0.44} & \textit{0.51} & \textit{0.23} & \textit{0.43} & \textit{0.45} & \textit{0.52} & \textit{0.43} & \textit{0.51} \\
DA17 & 0.80 & 0.70 & 0.77 & 0.42 & 0.68 & 0.70 & 0.77 & 0.67 & 0.77 \\
K17 & 0.55 & 0.44 & 0.50 & 0.28 & 0.44 & 0.45 & 0.51 & 0.43 & 0.50 \\
 \hline
\end{tabular}
\caption{This table is the same as \mbox{Table \ref{tb:integral_rates}}, with rates added for the CLLF LISA configuration. Merger and detection rate calculations per year are shown using the integral calculation (Equation \ref{eq:coalrate}). The evolution prescriptions are listed in the first column. The top row shows the ``no delays'' model: ND. The second row shows the ND-6 model, displayed in italics because it represents a subset of the ND model with $m_1,m_2\geq10^6M_\odot$. The final two rows show the DA17 \citep{Dosopoulou2017} and K17 \citep{PTA-illustris, Kelley2017b} models, respectively. The merger rate gives the rate of coalescences without considering LISA detectability. The remaining columns are labeled with the sensitivity curve used (PL or CLLF) to determine detectability. Additionally, detection rates are separated into signal types: All, Ins, MR, and BHS. ``All'' indicates reaching detection threshold using the entire signal. ``Ins'' and ``MR'' represent detection rates of inspiral signals and signals from the merger and ringdown, respectively. These categories are not independent. A single binary can add to the rate in both categories. ``BHS'' is the detection rate of MBH binaries where black hole spectroscopy is possible (see Equation \ref{eq:rhoGLRT}). We estimate our errors in these predictions to all fall below 0.01 according to equation 17 in \citet{Salcido2016}.}
\label{tb:integral_rates_all}
\end{table*}

 Here, we analyze the difference between the classic LISA sensitivity \citep{Larson2000} and the proposed LISA sensitivity \citep{LISAMissionProposal}. We do this purely to draw comparisons and understand how the measurement of MBH binaries has changed with the change in mission design. For this section, we refer to the Proposal sensitivity curve \citep{LISAMissionProposal} as ``PL.'' The classic LISA curve has an unrealistic low-frequency behavior following an $f^{-2}$ power law to infinitely low frequencies. To correct for this, we copy the low-frequency band edge behavior of PL, move it to lower strains, and spline it together with the classic LISA curve. We refer to this curve as ``CLLF.'' See \citet{Katz2018} for more information about this construction. Both PL and CLLF are shown in Figure \ref{fig:sensitivity_curves_and_signals2}. The basic difference between these two curves is better low-frequency performance exhibited by CLLF due to a longer armlength (5 million km).

The CLLF LISA configuration allows for observation of sources earlier in their inspiral as well as stronger overall measurements of this signal in the low-frequency regime. Additionally, the low-frequency difference can allow for detection of larger total mass sources \citep{Katz2018}. The difference between these two curves can be seen visually in Figure \ref{fig:sensitivity_curves_and_signals2} as the space between the two curves on the low-frequency end. For this paper, we study MBHs of masses greater than $\sim10^5M_\odot$. This mass regime radiates gravitational waves observable at frequencies below $10^{-3}$ Hz. Therefore, we want to focus our sensitivity analysis on the low-frequency band edge. \citet{Katz2018} show that the high mass range observable by LISA is strongly dependent on the low-frequency band edge behavior and can reach masses of $\sim10^9M_\odot$.

We also tested the ability of each detector configuration to observe partial signals as well as binaries earlier in their evolution during only their inspiral phase. Figure \ref{fig:time_to_merger} shows the detection rate of binaries versus their time-to-merger at the start of LISA observation. The binaries shown in this Figure will not reach their coalescence by the time LISA observing terminates. Therefore, the rates shown represent detections from only the inspiral portion of coalescence. Immediately to the right of the observation time, the rate drops off significantly with the loss of the merger and ringdown signals. Additionally, the inspiral signal is not observable for larger binaries, meaning LISA will not detect any of the larger mass systems if they do not merge in the observing window. At longer times before merger, the rate is further decreased as signals from lower mass systems fall below the noise. Therefore, in order for these detections to occur, they have to be lower in mass, sufficiently close in luminosity distance, and close enough in time-to-merger to be detectable. Here, we see a stark difference between the two detector configurations. The PL configuration almost drops off entirely to the right of the observation time. CLLF displays a different behavior: it shows a more gradual decrease in the detection rate with increasing start time. Detections at start times between 4 and 
7 years occur at rates between $\sim10^{-2}$ yr$^{-1}$ and 1 yr$^{-1}$. Therefore, while our magnitude of the overall rates is low, the detection rate of these inspiral-only sources can enhance the overall detection rate by $\sim20\%$ if the detector's low-frequency performance is closer to the classic LISA configuration. If lower mass binaries were included, this percentage would increase because at masses lower than those tested here, we enter a regime where the inspiral stage can be observed, but the merger and ringdown are no longer detectable.

All of our delay models have the same general behavior for both LISA configurations. Since ND-6 bottoms out at $10^6M_\odot$, the difference between ND-6 and the other models can be see as the low-mass ($<10^6M_\odot$) contribution added by the advanced extraction. Consequently, ND-6 does have a much steeper drop for CLLF because it does not have the low-mass support to boost detections of inpiral-only sources. Once again, the advanced extraction helps to establish a more complete LISA analysis. Similarly, due to the inability of K17 to maintain the low-mass systems, the K17 curve is below the DA17 and ND model curves at all start times.

\subsection{Monte Carlo Rate Results}

Table \ref{tb:monte_carlo_rates} shows our Monte Carlo results for 10000 sampled catalogs. We estimate the errors in our Monte Carlo results to be approximately $\sqrt{1/N}=\sqrt{1/10000}=1\%$. The rates for the measurement of the entire signal roughly match our integral calculations \mbox{(Table \ref{tb:integral_rates_all})}. Testing the inspiral-only results was useful in this setting because we can get a rate without assuming specific start times. We see that PL measures a negligble rate of inspiral-only sources for all of our timescale models. CLLF, on the other hand, shows that its low-frequency performance constitutes $\sim20\%$ of all detections for models that include the lower mass binaries (9\% of detections for ND-6). These systems had the lowest masses in our catalogs.

%%%%%%%%%%%%%%%%%%%%%%%%%%%%%%%%%%%%%%%%%%%%%%%%%%

%%%%%%%%%%%%%%%%% APPENDICES %%%%%%%%%%%%%%%%%%%%%

%%%%%%%%%%%%%%%%%%%%%%%%%%%%%%%%%%%%%%%%%%%%%%%%%%

% Don't change these lines
\bsp	% typesetting comment
\label{lastpage}
\end{document}